\documentclass[showpacs,aps,prd,preprint,reprint,superscriptaddress,nofootinbib,longbibliography]{revtex4-2}
\usepackage[dvipdfmx]{graphicx}
\usepackage{amsmath,amssymb,bm,color,longtable,mathrsfs,amsfonts,slashed,ulem}
\usepackage[T1]{fontenc}       % ←これが重要
\usepackage[utf8]{inputenc}    % ソースがUTF-8なら

\usepackage[colorlinks=true, pdfstartview=FitV, linkcolor=magenta,citecolor=blue, urlcolor=magenta,
bookmarks=true, bookmarksnumbered=true, breaklinks]{hyperref}

\allowdisplaybreaks

\newcommand \beq{\begin{eqnarray}}
\newcommand \eeq{\end{eqnarray}}

\newcommand{\Nc}{N_{\rm c}}
\newcommand{\Nf}{N_{\rm f}}

\newcommand{\lqcd}{\Lambda_{\rm QCD}}

\newcommand{\vp}{ {\bm p}}
\newcommand{\vq}{ {\bm q}}

\newcommand{\br}{ {\bm r} }

\newcommand{\vK}{{\bm K}}

\newcommand{\vx}{ \bm{ {x}} }

\newcommand{\la}{\langle}
\newcommand{\ra}{\rangle}

\newcommand{\calB}{\mathcal{B}}
\newcommand{\calG}{\mathcal{G}}

\newcommand{\calN}{\mathcal{N}}
\newcommand{\calM}{\mathcal{M}}
\newcommand{\calF}{\mathcal{F}}
\newcommand{\calK}{\mathcal{K}}

\newcommand{\calP}{\mathcal{P}}

\newcommand{\rmd}{\mathrm{d}}
\newcommand{\rmi}{\mathrm{i}}
\newcommand{\rme}{\mathrm{e}}

\begin{document}

% --- 右上に preprint number を強制配置 ---
\begin{minipage}{1.0\linewidth}
    \flushright
KEK-TH-2822
    % \\
%    J-PARC-TH-XXX
\end{minipage}
\vspace{-1cm} % タイトルとの余白を調整（必要に応じて）

\begin{flushright}
\end{flushright}

\title{Delineating neutral and charged mesons in magnetic fields
}

\author{Toru Kojo}
\email{ torukojo@post.kek.jp }
\affiliation{ Theory Center, IPNS, High Energy Accelerator Research Organization (KEK), 1-1 Oho, Tsukuba, Ibaraki, 305-0801, Japan }
\affiliation{ Graduate Institute for Advanced Studies, SOKENDAI, 1-1 Oho, Tsukuba, Ibaraki, 305-0801, Japan }

\author{Sakura Itatani}
\email{ itasaku@post.kek.jp }
\affiliation{ Theory Center, IPNS, High Energy Accelerator Research Organization (KEK), 1-1 Oho, Tsukuba, Ibaraki, 305-0801, Japan }
\affiliation{ Graduate Institute for Advanced Studies, SOKENDAI, 1-1 Oho, Tsukuba, Ibaraki, 305-0801, Japan }

\date{\today}

\begin{abstract}
We investigate the properties of neutral and charged mesons in magnetic fields, from weak-field to strong-field regimes.
To develop analytic insights, 
we employ a non-relativistic quark model with a confining potential of the harmonic oscillator type.
Short-range correlations, such as Coulomb and color-magnetic interactions, are treated as perturbations.
In particular, we focus on the magnetic field dependence of the relative and the center-of-mass motions.
The qualitative trends differ significantly between neutral and charged mesons:
for neutral mesons, the transverse momenta are conserved and continuous,
while charged mesons exhibit quantized transverse dynamics.
The Zeeman effects, arising from intrinsic spins and orbital angular momenta,
are carefully examined.
In particular, for charged mesons with spins $s\ge 1$, 
we discuss how the zero-point energy in the internal quark motion
cancels the Zeeman energy from the orbital angular momentum, ensuring the energetic stability of mesons with high spins.
The effectively reduced dimensionality of these mesons in the strong-field limit is also discussed.
\end{abstract}


\maketitle
\preprint{KEK-TH-2822}

%%%%%%%%%%%%%%%%%%%%%%
\section{Introduction}
%%%%%%%%%%%%%%%%%%%%%%

Quantum chromodynamics (QCD) in strong magnetic field regimes
has attracted significant attention
because of its importance in heavy-ion collisions \cite{Kharzeev:2007jp,Skokov:2009qp,Deng:2012pc,Fukushima:2008xe} and 
its potential applications to neutron star physics \cite{Duncan:1992hi,Chakrabarty:1996te,Ferrer:2005vd,Gyory:2022hnv,Ferrer:2021mpq,Ferrer:2015iop}, 
as well as its role as a critical testbed for theoretical concepts (for reviews, see, e.g. Refs.~\cite{Adhikari:2024bfa,Hattori:2016emy,Kharzeev:2015znc,Miransky:2015ava,Andersen:2014xxa,Iwasaki:2021nrz}).
Strong magnetic fields couple to quarks and
quantize their transverse dynamics,
which in turn affects the vacuum structure and the properties of hadrons \cite{Miransky:2002rp,Gusynin:1995nb,Endrodi:2013cs,Fukushima:2012kc,Kojo:2012js,Kojo:2013uua,Kojo:2014gha,Hattori:2015aki,Kojo:2021gvm,Arifi:2025ivt,Alford:2013jva,Fraga:2012ev,Fraga:2012fs,Ebert:2003yk,Shushpanov:1997sf,Cao:2021rwx,Braun:2014fua,Mueller:2015fka,Mueller:2014tea,Wang:2026xsm,Mei:2026xlj,Mei:2024rjg,Mei:2022dkd,Mao:2018dqe,Andreichikov:2018wrc,Andreichikov:2016ayj,Simonov:2016xaf,Moreira:2022dwo,Avancini:2016fgq,Ayala:2018zat,Avancini:2018svs,Ayala:2020dxs,Ayala:2023llp,Avancini:2022qcp}.
These phenomena have been extensively studied in lattice Monte-Carlo simulations \cite{Ding:2026qzu,Ding:2023bft,Ding:2021cwv,Ding:2020hxw,Endrodi:2019zrl,Bali:2017ian,Bali:2014kia,Bruckmann:2013oba,Bali:2012zg,Bali:2011qj,Endrodi:2015oba,DElia:2021tfb,Bonati:2016kxj,Bornyakov:2013eya,Hidaka:2012mz,Hattori:2019ijy},
covering aspects such as confinement/deconfinement transitions, chiral restoration, hadron spectra, and thermodynamics.
However, the physical interpretation of these lattice results is often not straightforward, 
triggering renewed interest in the fundamental aspects of QCD dynamics in extreme environments.

In this work we study mesonic states in magnetic fields, from weak $(eB \ll \lqcd^2)$ to strong field regimes $(eB \gg \lqcd^2)$
with $\lqcd \simeq 0.2$--0.3 GeV being a typical non-perturbative scale in QCD.
Both neutral and charged mesons are analyzed.
Our primary interest lies in how internal quark dynamics affects the overall behavior of mesons, 
which are frequently treated as point-like particles in effective models. 
By isolating the relevant effects one by one, 
we discuss how the parameters in hadronic effective field theories (EFT) 
should be modified in the presence of magnetic fields.
Specifically, in this paper we focus on the magnetic-field dependence of the meson masses and their transverse kinetic terms. 
For charged mesons, especially for states with spin and/or orbital angular momenta, 
the Zeeman terms tend to be cancelled by the zero-point energies associated with internal quark motion; this cancellation is one of the key composite-particle effects examined in this work.
The analysis of hadronic coupling constants will be presented in a subsequent paper.

Our primary goal is to develop analytic insights and establish a simple, intuitive physical picture.
(See Refs.~\cite{Coppola:2023mmq,GomezDumm:2023owj,Carlomagno:2022arc,Carlomagno:2022inu,Coppola:2019uyr} for comprehensive studies based on Nambu-Jona-Lasinio or quark-meson models.)
To this end, we adopt a simple non-relativistic quark model \cite{Isgur:1979be,DeRujula:1975qlm,Sakharov:1980ph,Zeldovich:1967rt}, 
which enables many quantities to be computed analytically \cite{Simonov:2016xaf,Simonov:2015yka}.
For applications to heavy quarkonia systems, see Refs.~\cite{Iwasaki:2021nrz,Alford:2013jva}.
In this work, we stick to the non-relativistic framework for several reasons.
While it is possible to extend this framework to a relativistic version \cite{Godfrey:1985xj,Capstick:1986ter}, 
such an extension introduces technical complexities that might distract us from developing a simple and intuitive picture.
Furthermore, for ground-state mesons, the large transverse kinetic energies at high $B$ (which typically challenge the validity of the non-relativistic framework)
are largely cancelled by Zeeman effects, as seen in relativistic frameworks (see, e.g., Ref.~\cite{Kojo:2013uua})\footnote{
% corrected
For example, for a free spin-1/2 elementary particle with charge $q>0$, 
the relativistic Landau energy may be written as
$E^2=m^2+p_z^2+qB (2n+1-\sigma_z)$.
For the lowest Landau level, the non-relativistic expansion gives
$E = m + \frac{p_z^2+qB(1-\sigma_z)}{2m}+\cdots$.
Thus the apparently large $O(B/m)$ contribution disappears for $\sigma_z=1$.
}. 
The remaining $B$-dependence is relatively weak, 
suggesting that a non-relativistic framework can still effectively capture the key properties of these low-lying states,
to a similar extent as non-relativistic models explain low-lying states at $B=0$.

This work builds upon and extends our previous investigations into this topic \cite{Hattori:2015aki,Kojo:2021gvm}. 
In this paper, we present a more comprehensive and conceptually refined analysis compared to our prior work, 
laying the groundwork for forthcoming studies that will encompass multi-quark systems and hadronic media. 
Specifically, we have significantly improved our treatment of the pseudo-momentum and guiding center coordinates, 
given more systematic and structured descriptions of charged mesons, 
and developed more physically motivated treatments of short-range correlations
which require special care in the limit of large magnetic fields.

This paper is structured as follows.
In Sec.~\ref{sec:preparation} we review the kinetic motion of charged particles in magnetic fields,
and in Sec.~\ref{sec:long_short} summarize our modeling of long- and short-range interactions,
and adjust the parameters to reproduce meson spectra at $B=0$.
In Sec.~\ref{sec:neutral_mesons} we study neutral meson spectra at finite $B$.
In Sec.~\ref{sec:charged_mesons} we analyze charged mesons.
Sec.~\ref{sec:summary} is devoted to summary.

%%%%%%%%%%%%%%%%%%%%%%
\section{Preparation: Kinetic motions of charged particles} \label{sec:preparation}
%%%%%%%%%%%%%%%%%%%%%%

%%%%%%%%
\begin{figure}[t]
\vspace{-.cm}
\begin{center}
\includegraphics[width=6.0 cm]{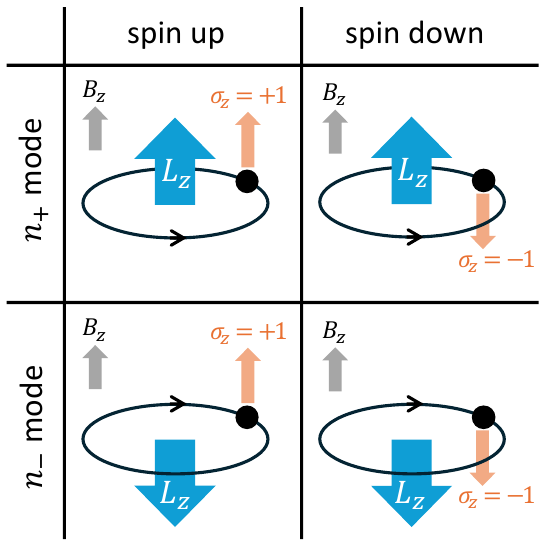}
\end{center}
\vspace{-.cm}
\caption{
Schematic illustration of the Zeeman splitting for orbital and spin angular momenta of a positively charged particle. 
The energy is lowered (raised) when the orbital angular momentum $L_z$ or the spin $\sigma_z$ 
is parallel (antiparallel) to the external magnetic field $B_z$. 
In the symmetric gauge, the $n_+$ mode corresponds to the orbital motion aligned with the magnetic field.
}
\label{fig:table_n_spin}
\end{figure}   
%%%%%%%%

In this section we summarize our setup,
reviewing the basics of the transverse motion in magnetic fields.
Throughout this paper, the quark mass refers to the constituent quark mass,
and we assume it to be $B$-independent.
This assumption is supported by the lattice results for the chiral condensate at $B \gg \lqcd^2$,
$\la \bar{q} q \ra \sim |eB| \la \bar{q} q \ra_{\rm 2D} \sim |eB| \lqcd$ \cite{Bali:2012zg},
where the overall factor $|eB|$ comes from the Landau degeneracy
and $\la \bar{q} q \ra_{\rm 2D}$ is computed from quark propagators with the dimensional reduction.
If the dynamically generated quark masses were sensitive to the magnetic field,
it would be very difficult to explain the linear scaling with respect to $B$ \cite{Kojo:2012js}.

%%%%%%%%%%%%%%%%%%%%%%
\subsection{Single charged particle } \label{sec:single}
%%%%%%%%%%%%%%%%%%%%%%

First we consider a single charged particle.
We consider a non-relativistic Hamiltonian
\beq
H_{\rm single} = m + \frac{\, \bm{\Pi}^2 \,}{\, 2m \,} - \bm{\mu} \cdot \bm{B} 
 \,,
\eeq
where $B$ is constant and applied in the $z$-direction.
The kinetic momentum and magnetic moment are
\beq
\bm{\Pi} = \vp - e \bm{A} \,,~~~~~ \bm{\mu} = \frac{\, e \,}{\, 2m \,} \bm{\sigma} \,,~~~~~\bm{B} = {\bm \nabla} \times \bm{A} \,,
\eeq
respectively.
(We take Land\'e g-factor to be $g=2$, treating the constituent quark as an elementary Dirac fermion.)

We define a pseudomomentum \cite{Alford:2013jva,Kojo:2021gvm}
\beq
\bm{\calK}
\equiv \bm{\Pi} + e \bm{B} \times \bm{r}
\equiv e \bm{B} \times \bm{X}
\,.
\eeq
This can be related to the {\it guiding center} coordinate
\beq
\bm{X} = - \frac{\, \bm{B} \times \bm{\calK} \,}{\, e B^2 \,} \,,~~~~~  B = | \bm{B}| \,,
\label{eq:guiding}
\eeq
which characterizes the center of the cyclotron motion.

The commutation rules among $\Pi$ and $\calK$ are
\beq
\big[ \Pi^x, \Pi^y \big] = \rmi e B_z = - \big[ \calK^x, \calK^y \big] \,,
%~~~~~~~~
%\big[ X_i, Y_j \big] = - \rmi \delta_{ij} \frac{1}{\, e_i B \,} \,,
\eeq
and all the other commutators are zero; 
in particular, $\big[ \Pi^\alpha, \calK^\beta \big] = 0 = \big[ \Pi^\alpha, X^\beta \big] $ for any sets of $(\alpha,\beta)$.

We note that $\Pi^x$ and $\Pi^y$ do not commute with $\bm{\Pi}^2$
and hence cannot be used as conserved quantities;
indeed, the Lorentz force keeps changing the direction of 
kinetic momentum $\bm{\Pi} \sim m \bm{v}$. 
Meanwhile, $\bm{\calK}$,
or the corresponding guiding center $\bm{X}$, commutes with $\bm{\Pi}^2$,
as the location of the center does not affect the energy spectrum.

For explicit computations it is convenient to choose a specific gauge.
Here we take the symmetric gauge
\beq
\bm{A} =  \frac{\,  \bm{B} \times \br \,}{2}  \,,
\eeq
with which
\beq
\bm{\Pi} = \vp - \frac{\, e \bm{B} \times \br \,}{2}  \,,
~~~~~~~
\bm{\calK} = \vp + \frac{\, e \bm{B} \times \br \,}{2}  \,.
\eeq
Writing $\bm{\Pi}^2 = p_z^2 + \bm{\Pi}_\perp^2$, we note
\beq
\bm{\Pi}_\perp^2 
= \vp_\perp^2 + \frac{\, (eB)^2 \,}{4} \br_\perp^2 - e B L_z \,.
\eeq
The first two terms correspond to a usual harmonic oscillator in two dimensions.
We first define the creation and annihilation operators for the $x$ component through
% corrected
\beq
\hspace{-0.5cm}
x = \frac{1}{\, \sqrt{|eB|} \,} \big( a_x + a_x^\dag \big) \,,
~~
p_x = -\rmi \frac{\, \sqrt{ |eB| }\,}{2} \big( a_x - a_x^\dag \big) \,,
\eeq
%
%%
%\beq
%\hspace{-0.5cm}
%x = \frac{1}{\, \sqrt{2|eB|} \,} \big( a_x + a_x^\dag \big) \,,
%~~
%p_x = -\rmi \sqrt{ \frac{\, |eB| \,}{2} } \big( a_x - a_x^\dag \big) \,,
%\eeq
%%
and also for the $y$ component in the same way.
Then we construct the creation and annihilation operators for 
$+$ and $-$ circular modes
\beq
a_\pm^\dag = \frac{1}{\, \sqrt{2} \,} \big( a_x^\dag \pm \rmi a_y^\dag \big) \,,
\eeq
with which
the angular momentum can be written as
\beq
L_z = a_+^\dag a_+ - a_-^\dag a_- \,.
\eeq
In our definition, the orbital magnetic moment points to the $z$ direction for the $+$ mode, see Fig.~\ref{fig:table_n_spin}.
Now, for later purposes we generalize the Hamiltonian for the transverse dynamics as
\beq
H_\perp 
= \frac{\, \vp_\perp^2 \,}{\, 2m \,} + \frac{\, m \omega_0^2 \,}{2} \br_\perp^2 - \Omega_0 L_z \,.
\eeq
For charged particles free from potentials,
$\omega_0 = |e B| /2m$ and $\Omega_0 = eB/2m$.
The spectrum for $H_\perp$ is
\beq
E^\perp_{n_r, \ell_z}
=   \omega_0 \big( 2 n_r + 1 \big)
+ \omega_0 \bigg( |\ell_z| - \frac{\, \Omega_0 \,}{\omega_0} \ell_z \bigg) 
\,,
\eeq
or equivalently,
\beq
E^\perp_{n_+,n_-}
=  \big( \omega_0 - \Omega_0 \big) n_+
+ \big( \omega_0 + \Omega_0 \big) n_- 
+ \omega_0
\,,
\eeq
where $2n_\pm = 2n_r + |\ell_z| \pm \ell_z$
with $n_r, n_\pm$ being non-negative integers and $\ell_z$ an integer.
The Zeeman energy associated with the orbital angular motion 
reduces the energy when its direction is aligned with that of magnetic fields,
while it costs more energy when the direction is opposite.
In the former case, the increase of $\ell_z$ costs more energy in the kinetic motion,
but the Zeeman coupling tends to cancel such energy cost.

In particular, for charged particles in the absence of external potential, 
we find $\omega_0 = \Omega_0$,
so that such cancellations are exact;
this degeneracy for $\ell_z \ge 0$ 
is nothing but the usual Landau degeneracy
for a given Landau level.
For $\omega_0 = \Omega_0$,
another way to understand this degeneracy is 
to interpret $n_+$ as a quantum number for the guiding center coordinates;
they should not affect the spectra because of the translational invariance.
When a charged particle is within an external potential or
interacts with another particle,
this degeneracy is lifted, as we discuss in the following sections.

%%%%%%%%%%%%%%%%%%%%%%
\subsection{ Two charged particles } \label{sec:two_charged}
%%%%%%%%%%%%%%%%%%%%%%

For two-body systems, the structure of our Hamiltonian is ($i,j$ are used as a particle label)
\begin{align}
H 
 = \sum_{j=1,2} \bigg[ m_j + \frac{\, \bm{\Pi}^2_j \,}{\, 2m_j \,} - \bm{\mu}_j \cdot \bm{B} \bigg] 
 + V (r_{12}) \,.
\end{align}
Here we included a potential which depends on the distance $r_{12} = |\br_1 - \br_2|$.

Unlike the single particle case, the potential commutes with neither $\bm{\calK}_1$ nor $\bm{\calK}_2$,
but commutes with the total pseudomomentum $ \bm{\calK}_1 + \bm{\calK}_2 \equiv \bm{\calK}_R$.
But in general we can label eigenstates of the Hamiltonian only by either $\calK_R^x$ or $\calK_R^y$,
since $\calK_R^x$ and $\calK_R^y$ do not commute in general,
\beq
\big[ \calK_R^x, \calK_R^y \big] = - \rmi e_R B_z  \,,~~~~ (e_R \equiv e_1 + e_2)\,.
\eeq
Clearly, the exception is the charge neutral case, $ e_R =0 $,
for which $\calK_R^x$ and $\calK_R^y$ can be used simultaneously
to label the eigenstates of the Hamiltonian.

%%%%%%%%%%%%%%%%%%%%%%
\section{ Long- and short-range interactions }  \label{sec:long_short}
%%%%%%%%%%%%%%%%%%%%%%

For two-body forces we separately treat
the long-range and short-range interactions.
Following the standard approach,
we treat the long-range confining potential at leading order
and the short-range part as perturbations.
In vacuum, it is essential to have the confining potential in our unperturbed Hamiltonian;
without it quarks would fly away and the perturbative framework never works.

%%%%%%%%%%%%%%%%%%%%%%
\subsection{ Confining potential }  \label{sec:conf_pot}
%%%%%%%%%%%%%%%%%%%%%%

For a confining potential, we use the harmonic oscillator 
which is standard in non-relativistic quark models,
\beq
V_{\rm conf} (r) = \lambda \br^2 
\,,~~~~~\lambda \sim \lqcd^3 \,.
\label{eq:confining_pot}
\eeq
Our motivation for using the harmonic oscillator comes simply from
its tractability in analytic computations.
More realistically, the confining potential should be linear at long distance
which is necessary to derive the Regge trajectories.
But in this work we restrict ourselves to low-lying states
so that the impact of this simplification should not be significant.

%%%%%%%%%%%%%%%%%%%%%%
\subsection{ Short-range potential } \label{sec:short_range}
%%%%%%%%%%%%%%%%%%%%%%

Next we turn to short-range interactions. 
As in usual quark models, we focus on the color-electric (Coulomb) 
and color-magnetic interaction.

%%%%%%%%%%%%%%%%%%%%%%
\subsubsection{ Color-electric interaction } \label{sec:VE}
%%%%%%%%%%%%%%%%%%%%%%

The color-electric interaction is
\beq
V_E(r) = -\frac{\, 4 \,}{\, 3 \,} \frac{\, \alpha_s \,}{\, r \,} \,.
\eeq
At strong magnetic fields, $eB \gg \lqcd^2$, 
actually the details of short-range interactions
become more important than the long-range interactions.
In fact, the model dependence largely appears from 
the treatment of the short-range correlations.

As we discuss later,
we evaluate the expectation values
of the following type,
\beq
\hspace{-0.5cm}
\la 1/r \ra
\sim
\Lambda_z \calB \int \frac{\, \rmd z \rmd^2 \br_\perp \,}{\, \sqrt{ z^2 + ( \br_\perp - \br_\perp^0)^2 } \,} 
\, \rme^{- \frac{\, \Lambda_z^2 \,}{2} z^2}  \rme^{- \frac{ \calB }{2} \br_\perp^2} 
\,,
\eeq
where we assume $\Lambda_z \sim \lqcd$ and $\calB^2 \sim B^2 + \lqcd^4$. 
The factor $\lqcd \calB$ and the Gaussian functions for $z$ and $r_\perp$ come from the
wavefunctions.
At large $B$, the range of $r_\perp$ is strongly limited to $\sim \calB^{-1/2}$.
Here we consider the case $\br_\perp^0 =0$.  
If we neglect the small $r_\perp$ in the denominator, 
we would encounter the integral 
\beq
 \calB \int_0^{\sim \Lambda_z^{-1}}
  \rmd z \int_0^{\sim \calB^{-1/2}} \!\! \frac{\, \rmd^2 \br_\perp \,}{\, z \,} 
\sim %\frac{1}{\, \calB \,}
 \int_0^{\sim \Lambda_z^{-1} } \frac{\, \rmd z \,}{\, z \,} \,,
\eeq
which diverges logarithmically because of the short distance singularity around $z \sim 0$.
For such small $z$, the small $1/\calB$ term in $r$ must be kept to cut off the UV contributions, leaving 
\beq
\big\la 1/r \big\ra
\sim
\Lambda_z \int_{\sim \calB^{-1/2} }^{\sim \Lambda_z^{-1} } \frac{\, \rmd z \,}{\, z \,}
\sim
\Lambda_z \ln \frac{\, \calB \,}{\, \Lambda_z^2 \,}
\,.
\eeq
This perturbation grows logarithmically with $\calB$;
at very large $\calB$ it would induce too much attraction between a quark and an antiquark,
driving the meson unstable.
Such instability, however, has not been found in lattice simulations.

Although this observation is based on a perturbative evaluation,
it is unlikely that non-perturbative treatment of $1/r$ forces cures the inconsistency with the lattice results;
with such improvement the wavefunction should have more concentration 
for smaller $r_\perp$ and the disagreement would become even worse.
Thus, a remedy for the instability, if it exists, 
should be found already at the evaluation in perturbative treatments.

A plausible scenario is that the running of coupling constants tempers
the growth of the attraction \cite{Simonov:2016xaf,Simonov:2015yka}.
At short distance of $\sim B^{-1/2}$,
the typical momentum transfer is large, and the corresponding value of $\alpha_s$ is small.
At large $B$, we use the one-loop running coupling ($\Nf=3$)
\beq
\alpha_s (Q^2) = \frac{\, 4\pi \,}{\, \beta_0 \ln \frac{\, Q^2 \,}{\, \lqcd^2 \,} \,} \,,
~~~~\beta_0 = 11 -2\Nf/3 \,,
\label{eq:alpha_s}
\eeq
taking $Q^2 \sim \calB$.
The logarithm in the denominator behaves as $\sim \ln (\calB/\lqcd^2)$,
which largely cancels the logarithm appearing in $\la 1/r \ra$.
Accordingly the $\la 1/r \ra$ does not grow indefinitely for $\calB\gg \lqcd^2$. 

In this paper we parametrize the effective momentum transfer 
as ($\Lambda_z^4 = 8\mu \lambda$)
%as ($\Lambda_z^2 = 8\mu \lambda$)
%
\beq
Q_B^2 \equiv m_1^2 + m_2^2 + \Lambda_z^2 + c_B eB \,,
\label{eq:QB2}
\eeq
and use the coupling $\alpha_s (Q_B^2)$ to evaluate the short-range interactions.
We try to cover the following situations:
(i) for larger quark masses, the wavefunction tends to be more compact so that $Q^2$ should be larger;
(ii) for $B \rightarrow 0$ and small $m_{1,2}$, the $Q^2$ should be characterized by the confining scale $\Lambda_z$,
(iii) for $B \gg \lqcd^2$, the momentum transfer should be large, $Q^2 \sim B$.

Here, the parameter $c_B$ is introduced because
our results for neutral mesons, which can have very compact structures in the transverse direction,
are sensitive to which $Q^2$ are used. 
In our numerical estimates we vary $c_B$ and show how the results are affected.

%%%%%%%%%%%%%%%%%%%%%%
\subsubsection{ Color-magnetic interaction }
%%%%%%%%%%%%%%%%%%%%%%

The traditional form of the color-magnetic interaction takes the form
\beq
V_M^{\rm trad} (r)
= \frac{\, 32 \pi \alpha_s \,}{\, 9 m_1 m_2 \,} \, \tilde{\delta} (\br) \, \bm{\sigma}_1 \cdot \bm{\sigma}_2 \,,
\label{eq:trad_color_mag}
\eeq
where the $\tilde{\delta} (\bm{r})$ can be the usual $\delta$-function 
or a strongly localized function such as $\sim \rme^{- \Lambda_{ {\rm UV} }^2 r^2 }$.
In momentum space, its Fourier transform leads to $\sim \rme^{-q^2/\Lambda_{ {\rm UV} }^2 }$, cutting off the momenta beyond $\Lambda_{\rm UV}$.

Since the expression~\eqref{eq:trad_color_mag} is arranged for the low energy regime with momentum transfer up to $\sim \Lambda_{\rm UV}$,
it is not clear whether the above parametrized form is reasonable also for the regime of $B \gg \Lambda_{\rm UV}$;
here the momentum transfer in the transverse direction can be as large as $B$, exceeding the cutoff scale.
Also there is a significant asymmetry between the $z$- and transverse directions. 

For this reason we examine the original relativistic expression which eventually leads to Eq.~\eqref{eq:trad_color_mag} in a certain limit.
We start from the spin dependent term in momentum space, 
\begin{align}
V_M^{\rm rela} 
& =
\frac{\, 16 \pi  \alpha_s \,}{\, 3 E_1 (\vp_1) E_2 (\vp_2)  \,} \calP_{ s} (\vq)
 % \frac{\, \bm{\sigma}_1\cdot \bm{\sigma}_2 - \big( \bm{q} \cdot \bm{\sigma}_1 \big) \big( \bm{q} \cdot \bm{\sigma}_2 \big)/\vq^2  \,}{\, E_1 (\vp_1) E_2 (\vp_2) \,}
\end{align}
with
\beq
\calP_s (\vq) \equiv \bm{\sigma}_1\cdot \bm{\sigma}_2 - \big( \bm{q} \cdot \bm{\sigma}_1 \big) \big( \bm{q} \cdot \bm{\sigma}_2 \big)/\vq^2 \,.
\eeq
Here the first factor arises from the vertices for a particle 1 and 2, $\bm{q}$ the momentum transfer, and $\bm{p}_{1,2}$ the momenta of the particle 1 and 2.
If $\bm{p}_{1,2}$ are small, one can make a replacement, $E_{1,2} \rightarrow m_{1,2}$, and take the spatial average $q_i q_j \rightarrow \delta_{ij} \vq^2/3$, 
with which a constant in momentum space yields the $\delta$-function in coordinate space.
Meanwhile, in the presence of large $B$, the momentum transfer and typical particle momenta are large, $\bm{p}_{1,2}^2 \sim \bm{q}^2 \gg m_{1,2}^2$.

In order to capture the trend of large momentum transfer while keeping the expression as simple as possible,
we consider the following compromised form
\begin{align}
V_M (\vq)
& \equiv
\frac{\, 16 \pi  \alpha_s \,}{\, 3 \,} \calF_s (m_1,m_2) \calP_s (\vq)
\end{align}
where we introduce a form factor ($\mu_{\rm IR}^{-2} \equiv m_1^{-2} + m_2^{-2}$)
\beq
\hspace{-0.4cm}
 \calF_s (m_1,m_2)
 \equiv 
 C_M \bigg[
 \frac{\, \rme^{ - \vq^2 /\mu_{\rm IR}^2 } \,}{\, m_1 m_2 \,}
 + \frac{\, 1 - \rme^{ - \vq^2 /\mu_{\rm IR}^2 } \,}{\, \vq^2 + c_m m_1 m_2 \,}
\bigg]
 \,.
 \label{eq:calF_s}
\eeq
The factor $C_M$ and $c_m$ are some factors of $O(1)$.
This form factor behaves as $\calF_s \rightarrow 1/m_1m_2$ for $\vq \rightarrow 0$ and
$\calF_s \rightarrow 1/\vq^2$ for $\vq \rightarrow \infty$, as requested.
The parameter $c_m$ is introduced 
just for the second term to vanish for $\vq \rightarrow0$.

In practice, we need to compute the following expectation value
\beq
\la V_M (r) \ra
= \int_\br |\psi(\br)|^2 \int_\vq \rme^{\rmi \vq \cdot (\br -\br_\perp^0) } V_M (\vq)
\,.
\eeq
For the ground state, the wavefunction is $|\psi(\br)|^2 \sim  \Lambda_z \calB \rme^{-\frac{\Lambda_z^2}{2} z^2}  \rme^{- \frac{\calB}{2} \br_\perp^2} $,
and we find
\begin{align}
\la V_M (r) \ra
& \sim  
\int_\vq V_M (\vq) ~
\rme^{- \frac{\, q_z^2 \,}{\, 2 \Lambda_z^2 \,} } \, \rme^{- \frac{\, \vq_\perp^2 \,}{\, 2\calB \,} -\rmi \bm{q}_\perp \cdot \bm{r}_\perp^0 }
\notag \\
& = \la V^s_{\rm I} \ra \bm{\sigma}_1 \cdot \bm{\sigma}_2
+ \la V^s_{\rm II} \ra \sigma^z_1 \sigma^z_2
\,,
\end{align}
where ($J={\rm I,II}$)
\beq
\la V_{J}^s \ra 
= \frac{\, 16 \pi  \alpha_s \,}{\, 3 \,}
\int_\vq 
\rme^{- \frac{\, q_z^2 \,}{\, 2 \Lambda_z^2 \,} } \, \rme^{- \frac{\, \vq_\perp^2 \,}{\, 2 \calB \,} -\rmi \bm{q}_\perp \cdot \bm{r}_\perp^0 }
\calF_s \calP_{J}\,,
\eeq
with 
\beq
\calP_{\rm I} 
= \frac{1}{\, 2 \,} \bigg( 1 + \frac{\, q_z^2 \,}{\, \vq^2 \,} \bigg)
\,,~~~~
\calP_{\rm II} 
= \frac{1}{\, 2 \,} \bigg( 1 - \frac{\, 3 q_z^2 \,}{\, \vq^2 \,} \bigg)
\,.
\label{eq:calP}
\eeq
The spin terms were split because the spatial wavefunctions are anisotropic.
At $B=0$ and $\br_\perp^0 =0$, one recovers the usual expression 
with $\calP_{\rm I} \rightarrow 2/3$ and $\calP_{\rm II} \rightarrow 0$.

The major modification in the present treatment
is the behavior at large $B$
which are improved by taking into account the scaling $E_{1,2} \sim |\vq|$ at large $\vq$.
The integration over $q_z$ yields a factor $\sim \lqcd$.
The integration for $\vq_\perp$ is divided to $|\vq_\perp | < m_{1,2}$ 
and $ m_{1,2} < |\vq_\perp | < \sqrt{B}$.
They yield
\beq
\la V_J^s \ra \sim c_1 \lqcd + c_2 \lqcd \ln \frac{\, B \,}{\, m_{1,2}^2 \,} \,,
\eeq
where $c_{1,2} \sim O(1)$.
As in the color-electric case,
we find the logarithmic enhancement for a large $B$,
which is largely cancelled by using the running coupling $\alpha_s (B)$.
Thanks to this, the perturbative corrections remain $\sim \lqcd$.

%%%%%%%%%%%%%%%%%%%%%%
\subsection{ Spectra at $B=0$ }  \label{sec:spectra_at_B0}
%%%%%%%%%%%%%%%%%%%%%%

Using the parametrization in the previous sections,
we adjust the parameters to reproduce
pseudoscalar mesons $(\pi, K)$
and vector mesons $(\rho, K_*, \phi)$.
The list of parameters is summarized in Table.~\ref{tab:meson_mass_fit_para}.
For this fitting procedure, only the first five parameters ($m_{u,d}, m_s, \lambda, \lqcd, C_M$) are important;
the parameter $c_m$ has a minor impact
while the term with $c_B$ becomes important only at large $B$.
When we optimize the parameters,
we first fix $\lqcd =0.250$ GeV for $\alpha_s$,
and then consider the range which we believe to be physically reasonable;
$m_{u,d} = 0.25$--$0.35$ GeV, $m_s = 0.40$--$0.55$ GeV,
$\lambda^{1/3} =0.10$--$0.25$ GeV, and $C_M = 0.1$--$3.0$.
Using $\lqcd =0.20$ GeV or $0.30$ GeV affects details of the optimization 
but does not change the overall trend.

In Table.~\ref{tab:meson_mass_fit_th_exp},
we examine various contributions to the spectra
by sequentially adding interactions;
starting with the sum of constituent quark masses,
we add confining interactions (+conf), Coulomb (+CE), and color-magnetic (+CM) interactions.
It is important to notice that the kinetic energy is large because
confinement forces quark wavefunction to be compact.
This kinetic energy is largely cancelled by the Coulomb interaction,
leaving the spectra close to the sum of constituent quark masses.
Finally, the color-magnetic interaction splits the spectra into 
the low- and high-lying modes.
In the following sections, we will delineate how these trends are affected by large magnetic fields.

\begin{table}[htbp]
  \centering
  \caption{Comparison of calculated and experimental meson masses (in GeV) with model parameters. 
  To reproduce the spectra of $(\pi, K, \rho, K_*, \phi)$,
  the first five parameters are important;
  meanwhile details of $c_m$ are not very important,
  and the term with $c_B$ is activated only at finite $B$.
  }
  \vspace{2mm}
  \begin{tabular}{clccccc}
    \hline \hline
   & &~~Value &~~~ Eq. &~~~~comment & & \\
    \hline
    & $m_{u,d}$ &~~~ 0.264\, [GeV] &~ & & & \\
    & $m_s$ &~~~ 0.425\, [GeV] &~ & & & \\
    & $\lambda^{1/3}$ &~~~ 0.170\, [GeV] &~~~\eqref{eq:confining_pot} &~~confining force & & \\
    & $\lqcd $ &~~~ 0.250\, [GeV] &~~~\eqref{eq:alpha_s} &~~running $\alpha_s$ & & \\
    & $C_M$ &~~~ 1.22 &~~~\eqref{eq:calF_s} &~~color-magnetic force & & \\
    & $c_m$ &~~~ 0.100 &~~~\eqref{eq:calF_s} &~~color-magnetic force  & & \\
    & $c_B $ &~~~ 0.1--1.0&~~~\eqref{eq:QB2} &~~~$Q^2$ at finite $B$ & & \\
     \hline \hline
      \end{tabular}
    \label{tab:meson_mass_fit_para}   
 \end{table}
 %%%
\begin{table}[htbp]
  \centering
  \caption{  Calculated masses include contributions from constituent masses ($m_1+m_2$), 
  confinement (+conf), Coulomb (+CE), and color-magnetic (+CM) interactions. 
  Experimental values are taken from the Particle Data Group (PDG).
  Model results for pseudoscalar meson $\eta_s = s\bar{s}$ are also shown. 
  }
 \begin{tabular}{lccccc}
    \hline \hline
     & $m_1 + m_2$ &~ +conf &~ +CE &\, Total (+CM) & Exp.(PDG) \\
    \hline
    $\pi$    & 0.528  &~ 0.937 &~ 0.610 & 0.147 & 0.140 \\
    $\rho$  & 0.528 &~ 0.937 &~ 0.610 & 0.764 & 0.775 \\
    $K$      & 0.689  &~ 1.06  &~ 0.805 & 0.486 & 0.494 \\
    $K^*$   & 0.689 &~ 1.06  &~ 0.805 & 0.911 & 0.892 \\
    $\eta_s$   & 0.850  &~ 1.17  &~ 0.946 & 0.698  & --- \\
    $\phi$   & 0.850  &~ 1.17  &~ 0.946 & 1.03  & 1.019 \\
    \hline \hline
  \end{tabular}
  \label{tab:meson_mass_fit_th_exp}  
\end{table}

%%%%%%%%%%%%%%%%%%%%%%
\section{Neutral mesons} \label{sec:neutral_mesons}
%%%%%%%%%%%%%%%%%%%%%%

In this section we discuss the properties of neutral mesons,
setting $q \equiv - e_1 = e_2$.
As mentioned, the total pseudomomentum is conserved.
Moreover $\calK_R^x$ and $\calK_R^y$ commute with each other.
It is instructive to express $\bm{\calK}_R$ in terms of the guiding centers,
\beq
\bm{\calK}_R 
= \sum_{j=1,2} e_j \bm{B} \times \bm{X}_j
= - \bm{B}_q \times \big( \bm{X}_1 - \bm{X}_2 \big) \,,
\eeq
where we introduced a notation $\bm{B}_q = q \bm{B}$.
This means that the distance between two guiding centers, $\bm{X}_1 - \bm{X}_2$, is conserved,
so that we expect the impact of interactions to appear to be
$\la V \ra \sim V(\bm{X}_1 - \bm{X}_2)$.

Below we choose coordinates as
\beq
\bm{R} = \frac{\, \sum_i m_i \bm{r}_i \,}{ M } \,,~~~~~ \bm{\rho} = \br_2 - \br_1 \,,
\eeq
with $M = m_1 + m_2 $.
The corresponding momenta are
\beq
\bm{P}_{R} = \sum_i \bm{p}_i \,,~~~~~~~
\bm{p}_{\rho} = - \frac{ m_2 }{\, M \,} \bm{p}_1 +  \frac{ m_1 }{\, M \,} \bm{p}_2 
 \,.
\eeq
We decompose our leading order Hamiltonian into the following pieces
\beq
H_0 = M + H_z + H_\perp + V_E + H_{\rm spin} \,,
\eeq
where
\begin{align}
H_z 
& = \frac{\, P_{R_z}^2 \,}{\, 2M \,}
+ \frac{\, (p_\rho^z)^2  \,}{\, 2\mu \,}
+ \lambda \rho_z^2
\,,
\notag \\
H_\perp 
& 
= \sum_{j=1,2} \frac{\, \big( \bm{\Pi}_j^\perp \big)^2 \,}{\, 2m_j \,}
+ \lambda \rho_\perp^2
\,,
\notag \\
V_E
& = -\frac{\, 4 \,}{\, 3 \,} \frac{\, \alpha_s \,}{\, \sqrt{ \rho_z^2 + \rho_\perp^2 } \,} 
\,,
\notag \\
H_{\rm spin} 
& 
= \frac{\, B_q \,}{\, 2 \,} \bigg( \frac{\, \sigma_1^z \,}{\, m_1 \,} - \frac{\, \sigma_2^z \,}{\, m_2 \,} \bigg)
+ V_M
 \,.
\end{align}
The most nontrivial part is the treatment of $H_\perp$,
and we will rewrite it shortly.
After deriving the eigenvalues of $H_0$, we construct perturbation theories
for short-range interactions.

%%%%%%%%%%%%%%%%%%%%%%
\subsection{Transverse motions}
%%%%%%%%%%%%%%%%%%%%%%

It is convenient to label eigenstates of the Hamiltonian by $\bm{K}_\perp$ such that
\beq
\bm{\calK}_R \Phi_{ \bm{K}_\perp } = \bm{K}_\perp  \Phi_{ \bm{K}_\perp } \,.
\eeq
Then, in the symmetric gauge 
\beq
\bm{\calK}_R = \bm{P}_R^\perp + \frac{\, \bm{B}_q \,}{2} \times \bm{\rho} 
\,.
\eeq
The solution of the eigenvalue equation for $\bm{\calK}_R$ is
\begin{align}
\Phi_{ \bm{K}_\perp } (\bm{R}_\perp, \bm{\rho} )
&= \exp \bigg[\, \rmi \bm{R}_\perp \cdot \bigg(  \bm{K}_\perp - \frac{\, \bm{B}_q \,}{2} \times \bm{\rho} \bigg) \, \bigg]
\varphi_{\bm{K}_\perp } ( \bm{\rho} ) 
\notag \\
&\equiv \rme^{\rmi \Theta_{\bm{K}_\perp} (\bm{R}_\perp,\, \bm{\rho} ) } 
\varphi_{\bm{K}_\perp } ( \bm{\rho} ) 
\,,
\end{align}
where the $R_\perp $-dependence is factorized.
Using this form, our eigenvalue equation for the Hamiltonian \cite{Alford:2013jva}
\beq
H_\perp \Phi_{ \bm{K}_\perp } (\bm{R}_\perp, \bm{\rho} )
= E_{ \bm{K}_\perp } \Phi_{ \bm{K}_\perp } (\bm{R}_\perp, \bm{\rho} )
\,,
\eeq
can be transformed into
\beq
H_\perp^{\rm eff,\, temp} \varphi_{\bm{K}_\perp } ( \bm{\rho} ) 
= E_{ \bm{K}_\perp } \varphi_{\bm{K}_\perp } ( \bm{\rho} ) \,,
\eeq
where $H_\perp^{\rm eff,\, temp} \equiv \rme^{ -\rmi \Theta_{ \bm{K}_\perp } } H_\perp \rme^{ \rmi \Theta_{ \bm{K}_\perp } }$.
The explicit form is
(reduced mass: $\mu^{-1} = m_1^{-1} + m_2^{-1}$)
\begin{align}
H_\perp^{\rm eff,\, temp} 
&=
 \frac{\, \big( \bm{p}^\perp_\rho \big)^2  \,}{\, 2\mu \,}
+ \bigg( \frac{\, B_q^2 \,}{\, 8\mu \,} + \lambda \bigg) \rho_\perp^2
+ \frac{\, K_\perp^2  \,}{\, 2M \,}
\notag \\
& \hspace{-1cm}
+ \frac{\, B_q \,}{\, 2 \,} \bigg( \frac{1}{\, m_1 \,} - \frac{1}{\, m_2 \,} \bigg) L_\rho^z
- \frac{\, \bm{B}_q \,}{M} \cdot ( \bm{\rho}_\perp \times \vK_\perp )
\,.
\end{align}
Here we have attached ``temp'' (temporal) since we will rewrite it shortly 
for shifted variables.

This expression can be misleading, 
since it would give impressions that 
the spectra depend on $\bm{K}_\perp$ even in the case of
non-interacting limit, $\lambda \rightarrow 0$;
for $\lambda=0$, there should be only two independent
charged particles which yield only quantized spectra.
The present confusing expression comes from 
the failure to eliminate all linear terms in dynamical variables.

In eliminating the linear terms, 
shifting $\bm{\rho}$ by a constant is insufficient, since the attempt to eliminate 
the $\bm{\rho} \times \bm{K}_\perp$ term
yields another linear term in $\vp_\rho^\perp$ from $L_\rho^z$.
We need to shift $\bm{\rho}_\perp$ and $\bm{p}_\rho^\perp$ at the same time,
$\bm{\rho}_\perp \rightarrow \bm{\rho}_\perp + \bm{\rho}_\perp^0$ 
and $\bm{p}_\rho^\perp \rightarrow \bm{p}_\rho^\perp + \bm{p}_0^\perp$.
This shift with constant terms does not change the commutation relations for $\bm{p}_\rho^\perp$ and $\bm{\rho}_\perp$.
Explicitly, the proper shifts turn out to be
\begin{align}
\bm{\rho}_\perp^0 
&= \frac{\, \vK \times \bm{B}_q \,}{\, B_q^2 + 2 M \lambda \,} 
\,,
\notag \\
\bm{p}_0^\perp
&= \frac{\, m_1 - m_2 \,}{\, 2M \,} 
 	\frac{\, B_q^2 \vK_\perp \,}{\, B_q^2 + 2 M \lambda \,}
 \,,
 \label{eq:shift}
\end{align}
which generate terms that cancel the $K_\perp^2/2M$ term.
One should also notice that these shift also affect the coordinates in
$V_E$ and $V_M$;
at large $K_\perp$, the $\rho_\perp^0$ serves as a cutoff for short-range interactions.

Now, with new variables the previous $H_\perp^{\rm eff, temp}$ is replaced with
\begin{align}
H_\perp^{\rm eff}
& 
=  \frac{\, \lambda K_\perp^2 \,}{\, B_q^2 + 2 M \lambda  \,}
+ \frac{\, (\bm{p}_\rho^\perp)^2  \,}{\, 2 \mu \,}
	+ \frac{\, \calB_q^2 \,}{\, 8\mu \,}  \bm{\rho}_\perp^2
\notag \\
&~~~
+ \frac{\,  \bm{B}_q \cdot \bm{L}_\rho \,}{\, 2 \,}
	\bigg( \frac{\, 1 \,}{\, m_1 \,} - \frac{\, 1 \,}{\, m_2 \,} \bigg)
\,,
\end{align}
where we introduced a notation
\beq
\calB_q^2 = B_q^2 + 8 \mu \lambda \,.
\label{eq:B_q}
\eeq
In the present expression, it is clear that 
$K_\perp^2$ term vanishes for $\lambda =0$; 
in this case, only discrete spectra appear,
as they should.

The qualitative meaning of the $K_\perp^2$ term becomes
clearer if we replace $K_\perp^2$ with the guiding center coordinates,
\beq
\frac{\, \lambda K_\perp^2 \,}{\, B_q^2 + 2 M \lambda  \,}
= \frac{\, \lambda \big( \bm{X}_1 - \bm{X}_2\big) ^2 \,}{\, 1 + 2 M \lambda/B_q^2  \,}
\,.
\label{eq:trans_kin_neu}
\eeq
At large $B$, this is nothing but the energy from confining potential
whose separation is specified by the conserved distance 
between the guiding centers for two particles.
Note also that the limit $B_q \rightarrow 0$ reproduces the standard $K_\perp^2/2M$ term.

The spectra take the form
\begin{align}
E^\perp_{n_r, \ell_z}
&
= \frac{\,  \lambda K_\perp^2 \,}{\, B_q^2 + 2 M \lambda  \,}
+ \frac{\, \calB_q \,}{\, 2 \mu \,} \big( 2 n_r + 1 \big)
\notag \\
&\hspace{+0.2cm}
+ \frac{\, \calB_q  \,}{\, 2 \,}  
\bigg[ \frac{\, |\ell_z| \,}{\,  \mu \,} 
	+ \ell_z \frac{\, B_q \,}{\, \calB_q \,} \bigg( \frac{\, 1 \,}{\, m_1 \,} - \frac{\, 1 \,}{\, m_2 \,} \bigg) 
	\bigg] \,,
\end{align}
%
%%
%\begin{align}
%E_{\vK_\perp}^{n_+,n_-}
%& 
%= \frac{\,  \lambda K_\perp^2 \,}{\, B_q^2 + 2 M \lambda  \,}
%+ \frac{\, \calB_q \,}{\, 2 \mu \,} \big( n_+ + n_- + 1 \big)
%\notag \\
%&~~~
%+ \frac{\, B_q \,}{\, 2 \,} \bigg( \frac{\, 1 \,}{\, m_1 \,} - \frac{\, 1 \,}{\, m_2 \,} \bigg) 
%\big( n_+ - n_- \big)
%\,,
%\end{align}
%%
or equivalently,
\begin{align}
E_{\vK_\perp}^{n_+,n_-} \!
& 
= \frac{\, \lambda K_\perp^2 \,}{\, B_q^2 + 2 M \lambda  \,}
+ \frac{\, \calB_q \,}{\, 2 \mu \,} 
\notag \\
&~~~ 
+ \frac{\, 1 \,}{\, 2 \,}  \bigg[ \frac{\, \calB_q \,}{\,  \mu \,} 
	+ B_q  \bigg( \frac{\, 1 \,}{\, m_1 \,} - \frac{\, 1 \,}{\, m_2 \,} \bigg) 
	\bigg]  n_+ 
\notag \\
&~~~
+ \frac{\, 1 \,}{\, 2 \,}  \bigg[ \frac{\, \calB_q \,}{\,  \mu \,} 
	+ B_q  \bigg( \frac{\, 1 \,}{\, m_2 \,} - \frac{\, 1 \,}{\, m_1 \,} \bigg) 
	\bigg]  n_-
\,.
\label{eq:shifts}
\end{align}
From this expression, it is clear that $\lambda \rightarrow 0$ yields
the spectra of two independent charged particles.
Note that the coefficients of $n_+$ and $n_-$ are manifestly positive definite,
and any internal excitations cost the energy of $\sim B$.
Finally we note that there is a zero-point energy $\calB_q/2\mu$
which, in the ground state, is largely cancelled by the Zeeman energy from 
the intrinsic spins.

%%%%%%%%%%%%%%%%%%%%%%
\subsection{Perturbative corrections and spin dependence}
%%%%%%%%%%%%%%%%%%%%%%

Using the eigenfunctions for $H_z + H_\perp$
with the quantum numbers $\bm{n} = (n_z, n_r, \ell_z)$,
we evaluate the short-range interactions
as the first order of perturbation,
\beq
\hspace{-0.5cm}
\la V_E \ra_{ \bm{n} } \,,
~~~~
\la V_M \ra_{ \bm{n} } 
= \la V^s_{\rm I} \ra_{ \bm{n} } ( \bm{\sigma}_1 \cdot \bm{\sigma}_2 )
+ \la V^s_{\rm II} \ra_{ \bm{n} } \sigma^z_1 \sigma^z_2 
\,.
\eeq
These expectation values are evaluated numerically.
The electric potential may be simply added to the eigen-energy from $H_z$ and $H_\perp$.
The magnetic part appears in the spin dependent part of the Hamiltonian
\begin{align}
\hspace{-0.4cm}
\la H_{\rm spin} \ra_{ \bm{n} } 
&= \frac{\, B_q \,}{\, 2 \,} \bigg( \frac{\, \sigma_1^z \,}{\, m_1 \,} - \frac{\, \sigma_2^z \,}{\, m_2 \,} \bigg)
\notag \\
&~~~
+ \la V^s_{\rm I} \ra_{ \bm{n} } ( \bm{\sigma}_1 \cdot \bm{\sigma}_2 )
+ \la V^s_{\rm II} \ra_{ \bm{n} } \sigma^z_1 \sigma^z_2 
\,.
\end{align}
Here $( \bm{\sigma}_1 \cdot \bm{\sigma}_2 ) $ can be diagonalized by a state with the total spin $S$, 
and the corresponding eigenvalue is $2 \big[ S(S+1) - 3/2 \big]$.
Meanwhile, the Zeeman energy is diagonal only for $|S=1, S_z=\pm 1 \ra$;
the Zeeman term converts $| S=0,S_z=0\ra$ into $|S=1,S_z=0\ra$ and vice versa.

For $|S=1,S_z=\pm 1\ra$, the spectra are
\beq
\hspace{-0.5cm}
\big(E_{S=1}^{S_z=\pm 1} \big)_{ \bm{n} }
= 
\pm \frac{\, B_q \,}{\, 2 \,} \bigg( \frac{\, 1 \,}{\, m_1 \,} - \frac{\, 1 \,}{\, m_2 \,} \bigg)
+ \la V^s_{\rm I} + V^s_{\rm II} \ra_{\bm{n}} \,.
\eeq
For $S_z=0$ states, we diagonalize the matrix
\beq
\begin{bmatrix}
~ - 3 \la V^s_{\rm I} \ra_{\bm{n}} - \la V^s_{\rm II} \ra_{ \bm{n} } ~&~ \frac{\, B_q \,}{\, 2\mu \,}~\\
~ \frac{\, B_q \,}{\, 2\mu \,} ~&~ \la V^s_{\rm I} \ra_{\bm{n}} - \la V^s_{\rm II} \ra_{ \bm{n} } ~
\end{bmatrix}
\eeq
and find the spectra ($\la V^s_{\rm I + II} \ra = \la V^s_{\rm I } + V^s_{\rm II}\ra$)
\beq
\hspace{-0.5cm}
\big(E_{S=\pm}^{S_z=0} \big)_{ \bm{n} }
= - \la V^s_{\rm I + II} \ra_{ \bm{n} }
\pm 
\sqrt{ 4 \la V^s_{\rm I} \ra_{ \bm{n} }^2 + \bigg( \frac{\, B_q \,}{\, 2\mu \,} \bigg)^2 \,}
\,.
\eeq
Among all these spectra, the $S_z=0$ state with the energy $\big(E_{-}^{S_z=0} \big)_{ \bm{n} }$
leads to the smallest energy.

%%%%%%%%%%%%%%%%%%%%%%
\subsection{Spatial wave functions}
%%%%%%%%%%%%%%%%%%%%%%

Short-range interactions are numerically evaluated
using the wavefunctions of the unperturbed Hamiltonian.
%Below we set $K_\perp =0$.
The spatial wavefunctions take the forms 
\beq
\Phi_{\bm{n}}^{\rm nt}  \big( z,\bm{\rho}_\perp \big)
= \psi_{n_z} (z) \phi_{\bm{n}_\perp }^{\rm nt}  ( \bm{\rho}_\perp )
\,,
\eeq
where each wavefunction is normalized to one.
The wavefunction in the $z$-direction is 
\beq
\psi_{n_z}(z)
= \calN_{n_z} H_{n_z} (z_\Lambda) \rme^{ - \frac{\, \Lambda^2 \,}{ 4 } z^2 }
\,,
\eeq
with $z_\Lambda = (2\mu \lambda)^{1/4} z \equiv \Lambda z/\sqrt{2}$, a normalization constant
$\calN_{n_z} = \big( \Lambda/2^{n_z} n_z!\sqrt{2\pi} \big)^{1/2}$,
and $H_{n_z}$ being the Hermite polynomials.
For the relative motion in the transverse direction, 
the unperturbed wavefunctions for neutral mesons
take the form 
\beq
\phi_{\bm{n}_\perp }^{\rm nt}  ( \bm{\rho}_\perp )
= \calN^{ {\rm nt} }_{ \bm{n}_\perp} \rme^{\rmi \ell_z \theta} \rho_{ \calB }^{| \ell_z | } 
	L_{n_\perp}^{| \ell_z |} \big(\rho_{ \calB }^2 \big)
	\rme^{- \frac{\, \calB_q  \,}{4} \rho_{ \perp}^2 } 
	\,,
\eeq
where $L_{n_\perp}^{| \ell_z |}$ is the associated Laguerre polynomials
and the normalization constant is 
$\calN^{ {\rm nt} }_{ \bm{n}_\perp} = [\, \calB_q  n_\perp!/ 2 \pi (n_\perp + |\ell_z| )! \,]^{1/2}$.
The effective transverse coordinate is $\rho_{\calB } = \rho_\perp (\calB_q/2)^{1/2}$
with the parameter $\calB_q$ which was given in Eq.~\eqref{eq:B_q}.

We are interested in the modes with $n_z = n_\perp = l_z =  0$
for which $H_{n_z=0}= 1$ and $L_{n_\perp=0}^{ |\ell_z| } = 1$.
We use the following wavefunction
\beq
\Phi_{\rm gs}^{\rm nt}  \big( z,\bm{\rho}_\perp \big)
= \calN_{n_z} \calN^{ {\rm nt} }_{ \bm{n}_\perp} 
	  \rme^{ - \frac{\, \Lambda^2 \,}{ 4 } z^2 } \rme^{- \frac{\, \calB_q \,}{4} \rho_{ \perp}^2  } 
\,,
\eeq
to evaluate the color-electric and magnetic interactions.
The analytic structure was examined in Sec.~\ref{sec:short_range}.
%Now we evaluate them numerically for the parameter sets used in a quark model.

Now we summarize the integral to be evaluated.
The expressions suitable for numerical evaluations are summarized in Appendix.~\ref{sec:int_formula}.
Including the shift $\bm{\rho}_\perp^0 \sim O(K_\perp/B)$ in Eq.~\eqref{eq:shift}, 
the color-electric energy is ($n_z=n_\perp =0$)
\beq
\hspace{-0.5cm}
\la V_E \ra
= - \frac{\, 4 \alpha_s \,}{3} |\calN_{n_z} |^2 |  \calN^{ {\rm nt} }_{ \bm{n}_\perp} |^2
\int_{\br} \frac{\, \rme^{- \frac{\, \Lambda^2 \,}{2} z^2 - \frac{\, \calB_q \,}{2} \bm{\rho}_\perp^2 } \,}{ \sqrt{ \rho_z^2 + (\bm{\rho}_\perp + \bm{\rho}_\perp^0)^2 } }
\,,
\eeq
whose qualitative trend was discussed in Sec.~\ref{sec:VE}.

The color-magnetic energy is
\beq
\la V_M \ra
 = \la V^s_{\rm I} \ra \bm{\sigma}_1 \cdot \bm{\sigma}_2
+ \la V_{\rm II}^s \ra \sigma^z_1 \sigma^z_2 \,,
\eeq
with ($J={\rm I, II}$)
\beq
\la V_J^s \ra
= \frac{\, 16 \pi  \alpha_s \,}{\, 3 \,}
\int_{\vq} 
\rme^{- \frac{ q_z^2 }{\, 2 \Lambda^2 \,} - \frac{\, \vq_\perp^2 \,}{\, 2 \calB_q \,} -\rmi \bm{q}_\perp \cdot \bm{r}_\perp^0 }  \calF_s \calP_{J}
 \,,
\eeq
where $\calF_s$ and $\calP_J$ are given in Eqs.~\eqref{eq:calF_s} and \eqref{eq:calP}, respectively.

%%%%%%%%%%%%%%%%%%%%%%
\subsection{Spectra }
%%%%%%%%%%%%%%%%%%%%%%

%%%%%%%%
\begin{figure}[t]
\vspace{-.1cm}
\begin{center}
\includegraphics[width=8.8 cm]{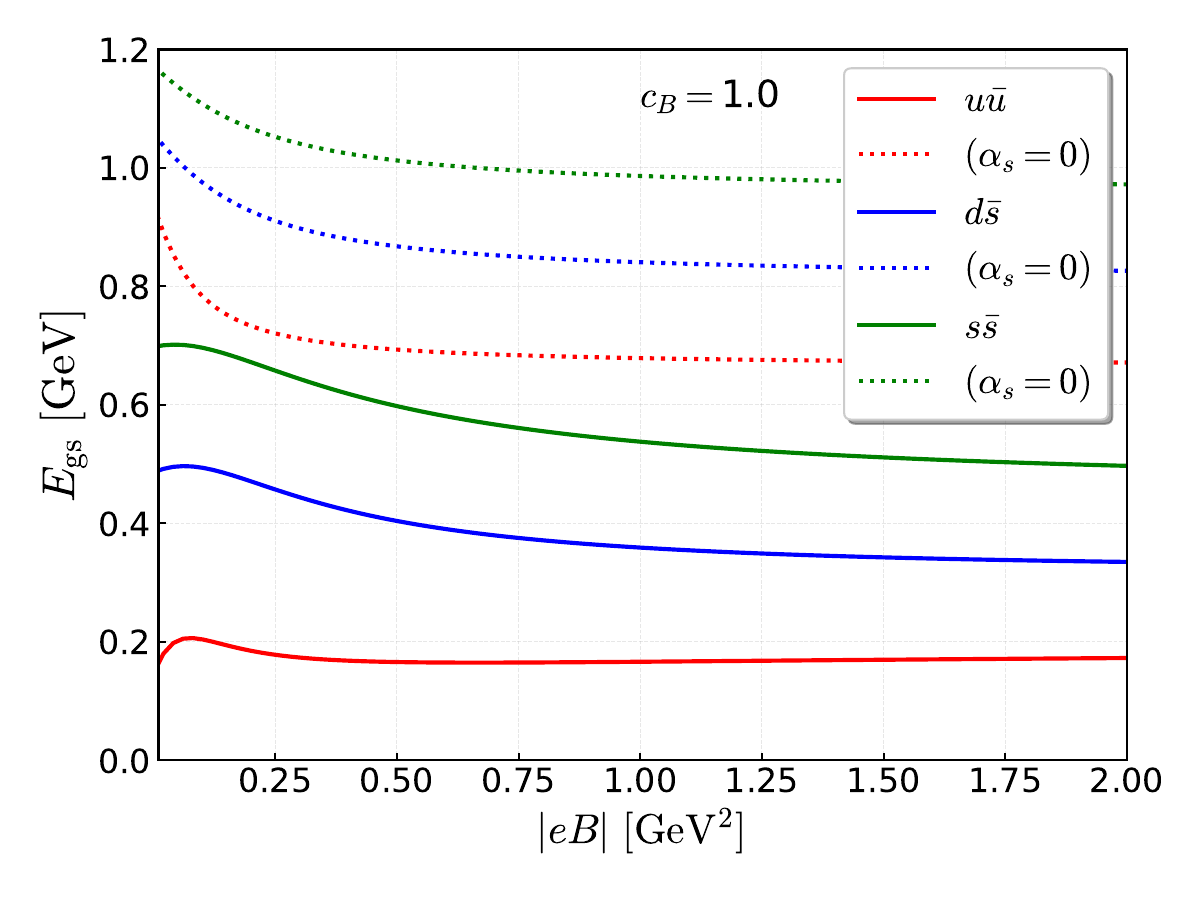}
\end{center}
\vspace{-.7cm}
\caption{Magnetic field dependence of the ground state energies for $u\bar{u}$, $d\bar{s}$, and $s\bar{s}$ mesons for $\vK=0$.
The solid and dotted lines represent calculations 
 with and without short-range interactions ($\alpha_s (Q_B) \neq 0$ and $\alpha_s = 0$).
 For $Q_B$, we use the expression with $c_B =1.0$.
}
\label{fig:uu_ds_alpha_on_off}
\end{figure}   
%%%%%%%%

%%%%%%%%
\begin{figure}[t]
\vspace{-.1cm}
\begin{center}
\includegraphics[width=8.8 cm]{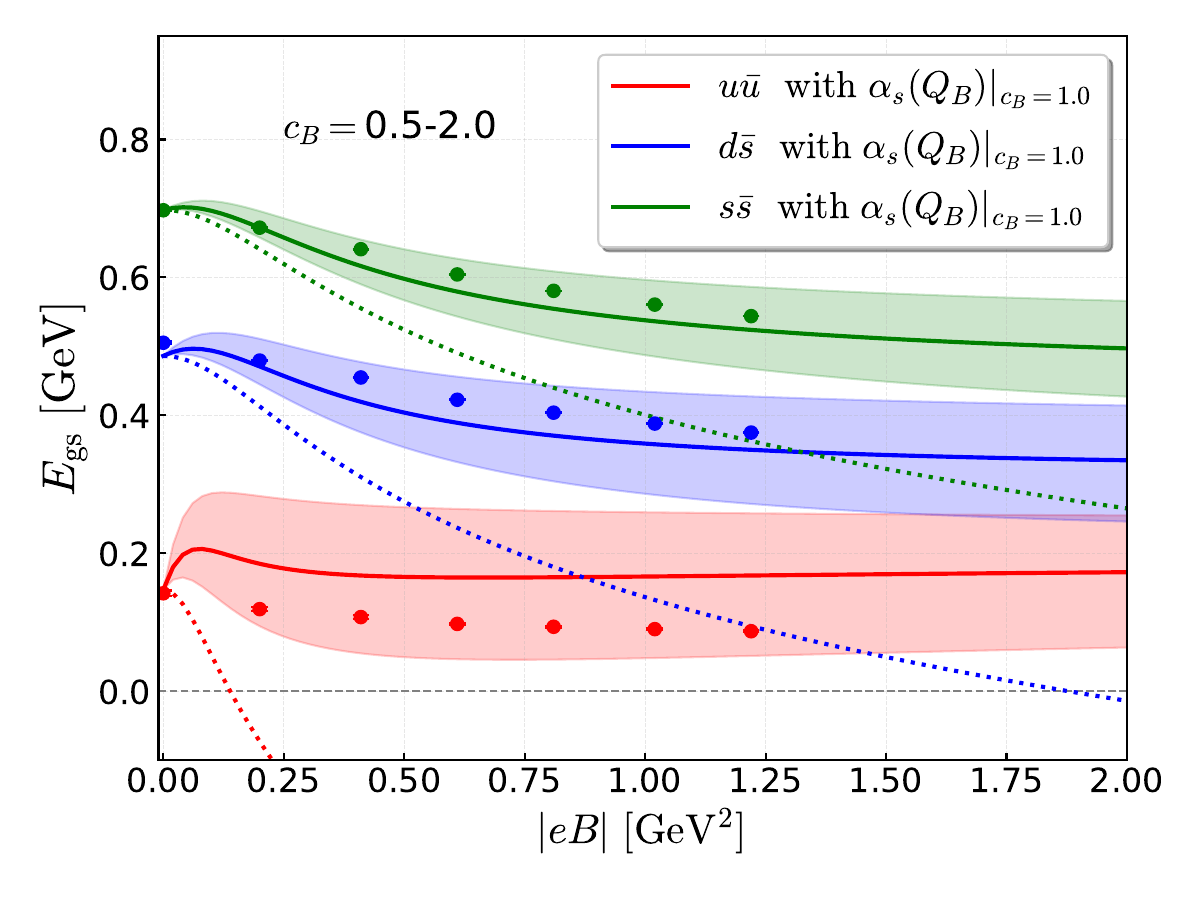}
\end{center}
\vspace{-.7cm}
\caption{Ground-state energy levels of $u\bar{u}$, $d\bar{s}$, and $s\bar{s}$ mesons as functions of $|eB|$,
with the running coupling $\alpha_s (Q_B)$ and $\vK=0$.
The colored bands represent the uncertainty originating from the scale parameter
$c_B \in [0.5,2.0]$, where the solid lines denote the central value, $c_B = 1.0$.
A smaller $c_B$ yields more mass reduction.
For comparison,
the results for a constant coupling (corresponding to $c_B=0$) are indicated with the dotted curves.
The data points are taken from the lattice QCD results reported in Ref.~\cite{Ding:2026qzu}.
}
\label{fig:uu_ds_alpha_varying}
\end{figure}   
%%%%%%%%

%%%%%%%%
\begin{figure}[t]
\vspace{-.1cm}
\begin{center}
\includegraphics[width=8.8 cm]{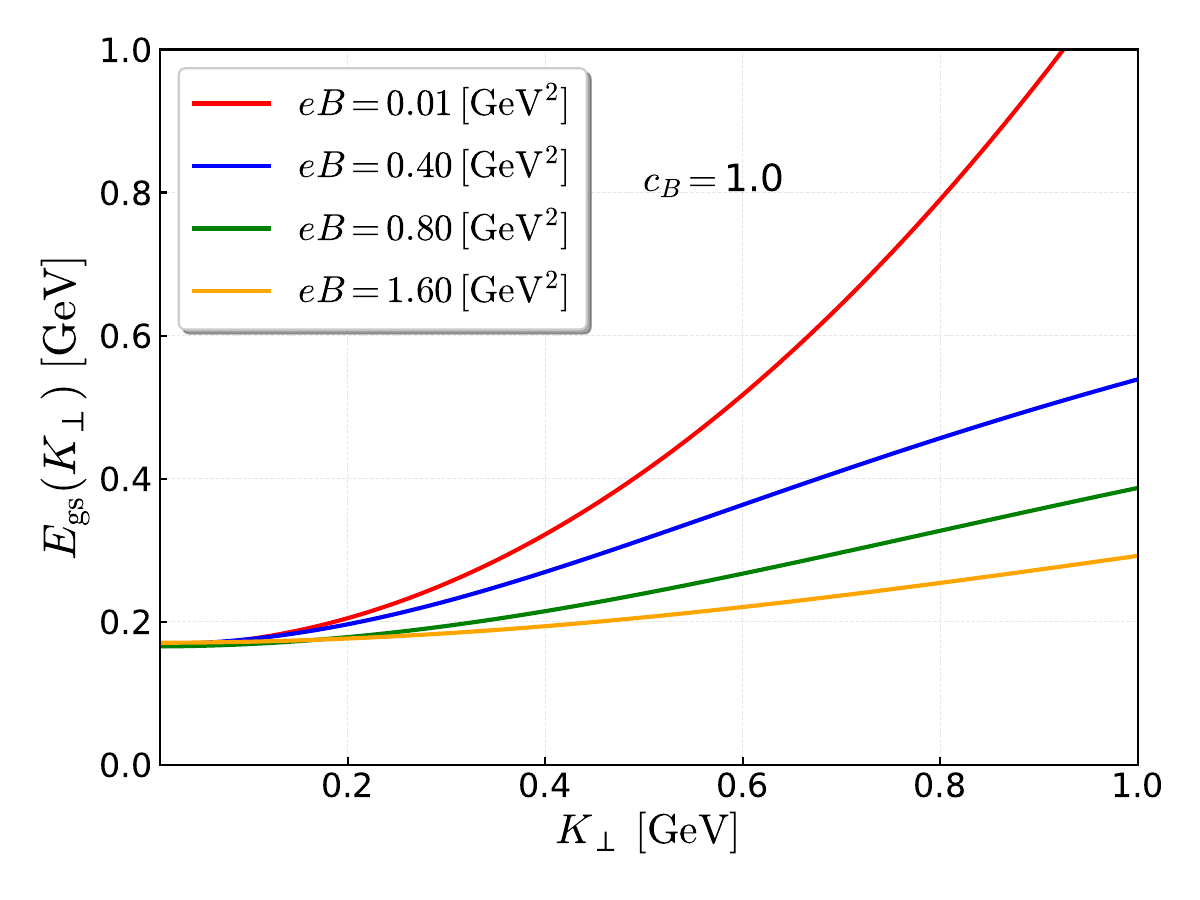}
\end{center}
\vspace{-.7cm}
\caption{The ground-state energies for the $u\bar{u}$ meson as functions of transverse momentum $\vK_\perp$,
for various $|eB|$.
At larger $B$, the spectra become more insensitive to the value of $K_\perp$.
$c_B=1.0$ is chosen.
}
\label{fig:uu_finite_K}
\end{figure}   
%%%%%%%%

Collecting the energies evaluated so far, 
the spectra for neutral mesons can be written as
\begin{align}
\hspace{-0.1cm}
 E_{ \bm{n} }^{S,S_z} (\vK) 
= M
+ E_{K_z}^{n_z}
+ E_{\vK_\perp}^{ \bm{n}_\perp }
+ \la V_E \ra_{\bm{n}}
+ \big(E_{S}^{S_z} \big)_{ \bm{n} }
\,.
\end{align}
We are particularly interested in the ground state energy 
given by the state $\bm{n}=0$, $S=-$, and $S_z=0$,
\begin{align}
\hspace{-0.1cm}
 E_{\rm gs} (\vK) 
 &= M + \la V_E \ra
+ \frac{\, K_z^2 \,}{\, 2 M \,}
+ \frac{\, \lambda K_\perp^2 \,}{\, B_q^2 + 2 M \lambda  \,}
+ \frac{1}{\, 2 \,} \sqrt{\frac{\, 2 \lambda \,}{\, \mu \,} } 
\notag \\
& \hspace{-0.3cm}
+ \frac{\, \calB_q \,}{\, 2 \mu \,} 
- \la V^s_{\rm I +II } \ra
- \sqrt{ 4 \la V_{\rm I}^s \ra^2 + \bigg( \frac{\, B_q \,}{\, 2\mu \,} \bigg)^2 \,}
\,.
\label{eq:neutral_gs}
\end{align}
The numerical evaluation is shown shortly.

We examine how the spectrum changes from weak to strong magnetic fields.
We set $\vK=0$.
At $B=0$,
\beq
\hspace{-0.5cm}
E_{\rm gs}^{B=0}  (0) 
= M + \la V_E \ra^{B=0} 
+ \frac{\, 3 \,}{\, 2 \,} \sqrt{\frac{\, 2 \lambda \,}{\, \mu \,} } 
- 3 \la V_{\rm I}^s \ra^{B=0} 
\,.
\eeq
At $B \gg \lqcd^2$,
\beq
\hspace{-0.5cm}
E_{\rm gs}^{B \gg \lqcd^2} (0) 
\simeq M + \la V_E \ra
+ \frac{\, 1 \,}{\, 2 \,} \sqrt{\frac{\, 2 \lambda \,}{\, \mu \,} } 
-  \la V_{\rm I + II }^{s} \ra
\,,
\eeq
where $O(1/B)$ terms are dropped.

The ground state spectra of $u\bar{u}$, $d\bar{s}$, and $s\bar{s}$ mesons for $\vK=0$ are shown in Fig.~\ref{fig:uu_ds_alpha_on_off}.
Except for a small enhancement at small $B$,
the ground state energies substantially decrease from their vacuum values.

To understand this behavior, it is instructive to switch off the short-range interactions by setting $\alpha_s \rightarrow 0$ (Fig.~\ref{fig:uu_ds_alpha_on_off}, dashed lines).
With increasing $B$, the zero-point energy of the transverse motion and the Zeeman energy largely cancel,
so the energy at large $B$ reduces as
\beq
\frac{\, 3 \,}{\, 2 \,} \sqrt{\frac{\, 2 \lambda \,}{\, \mu \,} } 
~ \rightarrow~
\frac{\, 1 \,}{\, 2 \,} \sqrt{\frac{\, 2 \lambda \,}{\, \mu \,} } + \frac{\, \calB_q - B_q \,}{\, 2 \mu \,} 
\,,
\eeq
where $ \calB_q - B_q \sim \mu \lambda/B_q$ is negligible at large $B$.
Hence the reduction at $\alpha_s=0$ may be understood
as the effective disappearance of the transverse kinetic energy.
For instance, the energy reduction in the large $B$ limit is 
$\sqrt{2\lambda/\mu} \simeq 0.272 $ (0.215) GeV for the $u\bar{u}$ ($s\bar{s}$) meson.

Now we switch on the short-range interactions, see
Figs.~\ref{fig:uu_ds_alpha_on_off} and \ref{fig:uu_ds_alpha_varying}
for $u\bar{u}$, $d\bar{s}$, and $s\bar{s}$ spectra.
Their impact sensitively depend on our choice of $Q^2$ in the running coupling $\alpha_s (Q^2)$,
as we have emphasized in Sec.~\ref{sec:short_range}.
If we neglect the running of $\alpha_s$ at a small transverse distance of $\sim B^{-1/2}$,
the energy reduction caused by the short-range interactions becomes too large (see the dotted lines in Fig.~\ref{fig:uu_ds_alpha_varying});
the spectrum in the $u\bar{u}$ channel becomes even unstable.
This too large reduction in the mass is tempered by using $\alpha_s$ with $Q^2 \sim B$.
While it is sensible to set $Q^2 \sim B$, we do not know precisely the numerical coefficient in front of $B$.
For this reason we vary the parameter $c_B$ for $Q_B^2$ in Eq.~\eqref{eq:QB2} for the range to be compatible with the lattice results.
In Fig.~\ref{fig:uu_ds_alpha_varying}, the band for the range $c_B \in [0.5, 2.0]$ is shown to display the impact of our choice on $Q_B$.
Lighter mesons are more sensitive to $Q_B$.

Finally we take a look at the ground-state energy for the $u\bar{u}$ meson at finite transverse momentum $\vK_\perp$
which is related to the distance between the two guiding centers as $|\bm{X}| \sim |\vK_\perp|/B_q$.
The energy shift at finite $\bm{\rho}_\perp$ originates from two competing factors:
(i) a reduction in the transverse center-of-mass kinetic energy [Eq.~\eqref{eq:trans_kin_neu}],
and (ii) an energy penalty arising from the weakened short-range attraction at finite $\bm{\rho}_\perp$.
The results are shown in Fig.~\ref{fig:uu_finite_K}.
As the magnetic field $B$ increases, 
the energy dependence on $K_\perp$ becomes less pronounced.
In the limit of $B\rightarrow \infty$, the energy becomes independent of $K_\perp$,
reflecting the effective dimensional reduction in the energy spectra.

%%%%%%%%%%%%%%%%%%%%%%
\section{Charged mesons} \label{sec:charged_mesons}
%%%%%%%%%%%%%%%%%%%%%%

%For charged mesons, the essential difference from the neutral case is that the total pseudomomentum components do not commute, leading to quantized transverse dynamics.
For charged mesons, the analyses of the spectra are much more involved.
 $\bm{\calK}_R$ is conserved as in the neutral meson case,
 but the $x$- and $y$-components do not commute, $[\calK_R^x, \calK_R^y] \neq 0$.
Below we choose the eigenvalue of $\calK_R^x$ to label the eigenstates of the Hamiltonian.
We look for the eigenstates of $\calK_R^x$,
\beq
\calK^x_R \Phi_{K_x} = K_x  \Phi_{K_x} \,.
\eeq
We expect a charged meson to circulate in a closed orbit
with which the spectra should be quantized.
Meanwhile, the eigenvalue equation for $\calK_R^x$ seems to yield a continuous eigenvalue $K_x$.
Combining these observations, 
we expect that the spectra become independent of $K_x$, making the spectra discrete.
We confirm this expectation shortly.

%%%%%%%%%%%%%%%%%%%%%%
\subsection{Transverse motions}
%%%%%%%%%%%%%%%%%%%%%%

For charged systems in coordinates $\bm{R}$ and $\br$, the pseudomomentum takes the form
(in the symmetric gauge)
\beq
\bm{\calK}_R 
%= \bm{P}_R + \frac{\, \bm{B} \,}{2} \times \sum_j e_j \br_j 
= \bm{P}_R + \frac{\, \bm{B} \,}{2} \times \big( e_R \bm{R} + e_\rho \bm{\rho} \big)
\,,
\eeq
where we write
\beq
%e_R = e_1 + e_2 \,,~~~~~
e_\rho = \frac{\, e_2 m_1 - e_1 m_2 \,}{ M } \,.
\eeq
We look for the eigenfunctions of $\calK_R^x$.
The equation to be solved is
\beq
\bigg[ - \rmi \frac{\partial}{\, \partial R_x \,} - \frac{B}{2} \big( e_R R_y + e_\rho \rho_y \big) \bigg] \Phi_{K_x} 
= K_x  \Phi_{K_x} \,.
\eeq
The solution takes the form
\begin{align}
&\Phi_{K_x} (\bm{R}, \bm{\rho}) 
\notag \\
&= \exp\bigg[ \rmi R_x \bigg( K_x + \frac{B}{2} \big(e_R R_y + e_\rho \rho_y \big) \bigg) \bigg] \varphi_{K_x} (R_y, \bm{\rho}) 
\notag \\
&
\equiv
\rme^{\rmi \Theta_c}  \varphi_{K_x} (R_y, \bm{\rho}) 
\,.
\end{align}
As we have done for the neutral meson case, we convert 
\beq
H_{c \perp} \Phi_{ K_x } (R_y, \bm{\rho} )
= E_{ K_x } \Phi_{ K_x } (R_y, \bm{\rho} )
\,,
\eeq
into
% corrected
\beq
H_{c \perp}^{\rm eff,\, temp} \varphi_{K_x } ( R_y, \bm{\rho} ) 
= E_{ K_x } \varphi_{ K_x } ( R_y, \bm{\rho} ) \,,
\eeq
%
%%
%\beq
%H_{c \perp}^{\rm eff,\, temp} \varphi_{K_x } ( \bm{\rho} ) 
%= E_{ K_x } \varphi_{ K_x } ( \bm{\rho} ) \,,
%\eeq
%%
where $H_{c \perp}^{\rm eff,\, temp} \equiv \rme^{ -\rmi \Theta_{ K_x } } H_{c \perp} \rme^{ \rmi \Theta_{ K_x } }$.
%
%The explicit form of $H_{c \perp}^{\rm eff,\, temp}$ requires a lengthy expression
%and we show it in the appendix.
%Below we only show how the $K_x$ dependence disappears from the spectra.

In the Hamiltonian, the $K_x$ comes out when we act $P_R^x$ on the phase $\Theta_c$.
We look at $\Pi_j^x$ ($j=1,2$)
\begin{align}
\rme^{-\rmi \Theta_c } \Pi_j^x \rme^{\rmi \Theta_c }
& = 
\frac{\, m_j \,}{M} \bigg[ K_x + B e_R R_y + P_R^x \bigg] 
+ \frac{B}{2} e_\rho \rho_y
\notag \\
&
- \eta_j \bigg[ 
p_\rho^x + \frac{B}{2} e_\rho R_y +  \frac{B}{2} \frac{\, \mu \,}{\, M \,} e_R \rho_y
\bigg] \,,
\end{align}
where $\eta_1 = -\eta_2 = 1$.
Now we consider a constant shift of the operators $R_y$ and $p_\rho^x$ as
\beq
R_y \rightarrow R_y - \frac{ K_x }{\, B e_R \,}\,,
~~~~
p_\rho^x \rightarrow p_\rho^x + \frac{\, e_\rho \,}{\, 2 e_R \,} K_x \,,
\eeq
with which the $K_x$ term disappears
\begin{align}
\rme^{-\rmi \Theta_c } \Pi_j^x \rme^{\rmi \Theta_c }
& \rightarrow
\frac{\, m_j \,}{M} \bigg[ B e_R R_y + P_R^x \bigg] 
+ \frac{B}{2} e_\rho \rho_y
\notag \\
&~~~
- \eta_j \bigg[ 
p_\rho^x + \frac{B}{2} e_\rho R_y +  \frac{B}{2} \frac{\,  \mu \,}{\, M \,} e_R \rho_y
\bigg] \,.
\end{align}
The other terms in the Hamiltonian do not contain $K_x$ terms
and are unaffected by the above constant shifts.
Hence we verified that the spectra are independent of the value of $K_x$.
For this reason, below we no longer attach the subscript $K_x$ 
to the energy and states.

After some calculations including the above-mentioned shifts, we arrive at
\begin{widetext}
\begin{align}
H_\perp^{\rm eff} 
&=  \frac{\, (P_R^y)^2 \,}{\, 2M \,}  
 + \bigg[ \frac{\, ( Be_R )^2 \,}{\, 2M \,} + \frac{\, ( B e_\rho)^2 \,}{\, 8\mu \,} \bigg] R_y^2
 + \frac{\, (\vp_\rho^\perp)^2 \,}{\, 2 \mu \,} 
 + \bigg[ \frac{\, \mu (  B e_R)^2 \,}{\, 8M^2 \,} 
 	+ \lambda
 	+ \frac{ (B e_\rho)^2 }{\, 8 \mu \,} 
	-  \frac{\,  \mu B^2 e_\rho e_R \,}{\, 4M \,} \bigg( \frac{1}{\, m_1 \,} -  \frac{1}{\, m_2 \,} \bigg)  
	\bigg] \bm{\rho}_\perp^2		
\notag \\
&~~~ 
+ \bigg[ - \frac{\, B e_R \,}{\, 2M \,}  + \frac{\, B e_\rho \,}{\, 2 \,}  \bigg( \frac{1}{\, m_1 \,} -  \frac{1}{\, m_2 \,} \bigg) \bigg] L_\rho^z 
\notag \\
&~~~ 
- \frac{\, 3 (B e_\rho)^2 \,}{\, 8 M \,} \rho_x^2
+ \bigg[ \frac{\, 5 B^2 e_R e_\rho \,}{\, 4M \,} - \frac{\, (B e_\rho)^2 \,}{\, 4 \,} \bigg( \frac{1}{\, m_1 \,} -  \frac{1}{\, m_2 \,} \bigg)   \bigg] \rho_y R_y 
+ \frac{\, B e_\rho \,}{\, 2 \mu \,} R_y p_\rho^x
- \frac{\, B e_\rho \,}{\, 2M \,} \rho_x P_R^y 
 \,.
\end{align}
\end{widetext}
The terms in the first two lines are axially symmetric and can be analytically diagonalized.
The complication comes from the third line which is not axially symmetric.
The full Hamiltonian contains anisotropic terms and the coupling between
the center-of-mass and internal motion.
In principle one can diagonalize analytically but the expression is too complicated.
We first consider the $e_\rho=0$ case for which we can derive the spectra analytically.
Then, we derive the full spectra numerically.

%%%%%%%%%%%%%%%%%%%%%%
\subsection{Analytic insights: $e_\rho =0$ case} \label{sec:e_rho=0}
%%%%%%%%%%%%%%%%%%%%%%

The $e_\rho=0$ case corresponds to $m_1=m_2 = M/2$ and $e_1=e_2=e_R/2$,
which are not realized in physical mesonic systems.
(For diquarks made of identical quarks, this condition can be satisfied \cite{Cao:2025rop}.)
But it is still useful to study this case as a baseline to diagonose a more general $e_\rho \neq 0$ case.

The energy spectra take the form
\begin{align}
\hspace{-0.1cm}
 E_{n_R, \bm{n}_c }^{S,S_z} (K_z)
= M
+ E_{K_z}^{n_z}
+ E_{n_R}^{ \bm{n}_\perp }
+ \la V_E \ra_{\bm{n}_c }
+ \big(E_{S}^{S_z} \big)_{ \bm{n}_c }
\,.
\end{align}
The short-range potentials are evaluated 
using the eigenfunctions of $H_0^\perp$.
The Hamiltonian at $e_\rho = 0$ is
\begin{align}
H^\perp_0 
&=
 \frac{\, (P_R^y)^2 \,}{\, 2M \,}  
 + \frac{\, ( B e_R )^2 \,}{\, 2M \,} R_y^2
 \notag \\
 &
 + \frac{\, 2 (\vp_\rho^\perp)^2 \,}{\, M \,} 
+ \bigg[ \frac{\, (B e_R)^2 \,}{\, 32 M \,} + \lambda \, \bigg] \bm{\rho}_\perp^2  
- \frac{\, B e_R \,}{\, 2M \,} L_\rho^z 
\,,
\end{align}
which yields the following spectrum
\begin{align}
\big( E_{n_R}^{ \bm{n}_\perp } \big)_0
& =  \frac{\, B e_R  \,}{\, 2M \,} \big( 2 n_R + 1 \big) 
\notag \\
&~
 + \frac{\, \calB_R  \,}{\, 2M \,} ( 2n_r + |\ell_z| + 1 )
- \frac{\, B e_R \,}{\, 2M \,} \ell_z 
\,,
\end{align}
or equivalently,
\begin{align}
\big( E_{n_R}^{ \bm{n}_\perp } \big)_0
& =  \frac{\, B e_R  \,}{\, 2M \,} \big( 2 n_R + 1 \big) 
\notag \\
&
 + \frac{\, \calB_R  \,}{\, 2M \,} ( n_+ + n_- + 1 )
- \frac{\, B e_R \,}{\, 2M \,} ( n_+ - n_- ) 
\,,
\end{align}
where 
\beq
\calB_R = \sqrt{ (B e_R)^2 + 32 M \lambda \,}
\,.
\label{eq:B_R}
\eeq
The angular momentum in a meson
tends to cancel the zero-point energy from the kinetic motion.
At large $B$, states with $\ell_z \ge 0$ are almost degenerate
with a small splitting due to $\calB_R - B e_R \sim \lambda M/B$.
Thus, at large $B$, there are many low energy excitations with $\ell_z \ge 0$.

There are also contributions from the spin-dependent term $\big(E_{S}^{S_z} \big)_{ \bm{n}_c }$.
The spin term is diagonalized by $(S,S_z)$,
\begin{align}
%\hspace{-0.5cm}
\big(E_{S}^{S_z} \big)_{ \bm{n}_c, 0 }
& = - \frac{\, B e_R \,}{\,  M \,} S_z 
+ \la V^s_{\rm I} \ra_{ \bm{n}_c } \big[\, 2 S(S+1) - 3 \, \big] 
\notag \\
&\hspace{0.5cm}
+ 4 \la V^s_{\rm II} \ra_{ \bm{n}_c } S^z_1 S^z_2 \,.
\end{align}
At large $B$, the ground state is given by the $S=S_z=1$ state
in which the Zeeman energy 
exactly cancels the zero-point energy from the center-of-mass motion
and almost completely cancels that from the relative motion.

The low-lying spectra which are insensitive to $B$ are given 
by $n_R=n_r=0$ modes with $\ell_z \ge 0$,
\begin{align}
\hspace{-0.5cm}
E_{\ell_z \ge 0}^{n_R=n_r=0}   
& = M 
+ E_{K_z=0}^{n_z=0}
+ \big\la V_E 
+ V_{\rm I+II}^s \big\ra_{n_r=0, \ell_z}   
\notag \\
&~~
 + \frac{\, \calB_R - B e_R \,}{\,  2M \,} \big( 1 + |\ell_z| \big) 
\,.
\label{eq:ch_gs_0}
\end{align}
The main findings from this special setup for large $B$ are:
(i) for $\ell_z > 0$, the orbital kinetic energy and Zeeman energy tends to 
cancel, leaving small excitation energies of $\sim \lambda /B$
which vanish for non-interacting limit, $\lambda \rightarrow 0$;
(ii) the zero-point energy of the center-of-mass motion 
cancels exactly with the half of the Zeeman energy;
(iii) the zero-point energy of the relative-motion
largely cancels with the half of the Zeeman energy;
(iv) in the limit of $\lambda \rightarrow 0$,
the energy levels of $\ell_z \ge 0$ are all degenerate.

%%%%%%%%%%%%%%%%%%%%%%
\subsection{Perturbative expansion with respect to $e_\rho$} \label{sec:pert_erho}
%%%%%%%%%%%%%%%%%%%%%%

For general cases of $e_1 \neq e_2$ and $m_1 \neq m_2$,
the expressions become complicated.
Still, if $\delta e = e_1 - e_2$ and $\delta m = m_1 - m_2$
can be regarded as small,
one can analytically compute corrections to the $e_\rho =0$ results.
The $e_\rho$ may be expressed as
\beq
e_\rho = \frac{\, e_R \delta m \,}{\, 2M \,} -  \frac{\, \delta e \,}{\, 2 \,} \,,
\eeq
which is the sum of $\delta e$ and $\delta m$ corrections.
In the light quark sector, the former is a good approximation,
and the latter may be regarded as the $1/\Nc$ corrections,
since $e_{u,d} = ( \pm 1 + 1/\Nc )/2$;
for charged combinations ($u, \bar{d}$) and ($d, \bar{u}$), 
the charge difference is $\delta e = e/\Nc$.

Computing to the second order of these perturbations, 
we find the correction to the ground state energy at large $B$, with the quanta $n_R=n_r=\ell_z=0$ and $S=S_z=1$, is
(for details, see Appendix.~\ref{sec:pert_details})
\begin{align}
\hspace{-0.1cm}
 \delta E_{\ell_z = 0}^{n_R=n_r=0} 
 &= - \frac{\, ( B e_\rho )^2 \,}{\, 2M \,} 
\bigg[
\frac{\, \big( \frac{\, \xi \,}{2} - \frac{\, 7 \,}{\, 2 \xi \,}   \big)^2 \,}
	{\,  \calB_R + B e_R \,}
+
\frac{\, \big(  \frac{\, \xi \,}{2} + \frac{\, 3 \,}{\, 2 \xi \,}  \big)^2 \,}
	{\,  \calB_R + 3 B e_R \,}
\bigg]		
\notag \\
& ~~~
 + \frac{\, B e_\rho^2  \,}{\, 4M e_R \,} 
 	\frac{\, \xi^2 + 10 \,} {\, \xi^2 \,}
+ \frac{\, 8 \delta m^2 \lambda  \,}{\, M^2 B e_R \xi^2 \,} 
\,,
\label{eq:ch_gs_2}
\end{align}
where $\xi = \big[ 1 + 32 M\lambda/(B e_R)^2 \big]^{1/4}$.
In the limit of $B\rightarrow \infty$, the scaling of parameters is
$\calB_R \rightarrow B e_R$, $\xi \rightarrow 1$.
In this limit the terms of $O(B)$ cancel, leaving
\beq
  \delta E_{\ell_z = 0}^{n_R=n_r=0}  ~\sim~ O(\lambda/B) ~~~~~~(B\rightarrow \infty) \,.
\eeq
Hence the insensitivity of the ground-state energy to $B$, as seen for the $e_\rho =0$ case,
remains valid to $O(\delta e^2, \delta m^2)$.
This conclusion holds also for $\ell_z >0$ with $n_R=n_r=0$.
This result indicates that
for general cases with $e_\rho \neq 0$,
the overall trend of the spectra remains similar as the $e_\rho =0$ cases.

%%%%%%%%%%%%%%%%%%%%%%
\subsection{General spectra for $e_\rho \neq 0$}
%%%%%%%%%%%%%%%%%%%%%%

The full Hamiltonian for the transverse kinetic term takes a complicated form
and we derive the spectra numerically.
What we need is the eigenfrequencies and the corresponding wavefunctions which
are utilized to compute the expectation values of the short-range interactions.

%%%%%%%%%%%%%%%%%%%%%%
\subsubsection{Eigenfrequencies}
%%%%%%%%%%%%%%%%%%%%%%

First we define 
$X = ( \rho_x, \rho_y, R_y, p_{\rho}^x, p_{\rho}^y, P_R^y )$
and write the Hamiltonian as
\beq
H_\perp^{\rm eff} 
= \frac{1}{\, 2 \,} X_i A_{ij} X_j  \,, ~~~~~A_{ij} = A_{ji} \,.
\eeq
The commutation relations for $X$'s are
\beq
[X_i, X_j] = \rmi J_{ij} \,,~~~~~ 
J = 
\begin{bmatrix} 
0 & I_{\bf 3 } ~ \\
~ -I_{\bf 3 } & 0 
\end{bmatrix}
\,,
\eeq
where $J$ is the fundamental symplectic matrix.
Our goal is to convert the Hamiltonian into the form
\beq
H_\perp^{\rm eff} 
= \sum_{\alpha=1}^3 \omega_\alpha \bigg( b^\dag_\alpha b_\alpha + \frac{1}{\, 2 \,} \bigg) \,,
\label{eq:HO_b}
\eeq
with the constraint $[b_\alpha, b_\beta^\dag] = \delta_{\alpha \beta}$,
and find the relation between $X$ and $(b,b^\dag)$,
\beq
X_i = T_{i \alpha} \tilde{b}_\alpha \,,~~~~~~~\tilde{b} = ( b_1, b_2,b_3, b_1^\dag, b_2^\dag, b_3^\dag) \,.
\label{eq:Tb}
\eeq
We can also express the commutation relation as $[ \tilde{b}_\alpha, \tilde{b}_\beta ] = J_{\alpha \beta}$.
This imposes the condition $\rmi J = T J T^{T}$
% $J = T J T^{T}$ corrected
which is satisfied in our ordering of operators in $\tilde{b}$.
The rest is the normalization of overall coefficients.

To find the spectra, we examine the Heisenberg equation
\beq
 \dot{X}_k = \rmi \big[ H_\perp^{\rm eff}, X_k \big] = ( JA)_{kj} X_j \,.
\eeq
We consider a linear transformation
%We consider a unitary transformation
$X_k = V_{k \alpha} \tilde{X}_\alpha $ with $V^{-1} (JA) V = {\rm diag} ( \lambda_1, \lambda_2, \cdots, \lambda_6)$,
\beq
 \dot{ \tilde{X} }_\alpha
 = \lambda_\alpha \tilde{X}_\alpha 
 ~\rightarrow~
 \tilde{X}_\alpha (t) = \rme^{ \lambda_\alpha t } \tilde{X}_\alpha (0) 
 \,.
 \label{eq:tilX_t_evo}
\eeq
We note that, if some component has the eigenvalue $\lambda_\alpha$, there is also a component having the eigenvalue $-\lambda_\alpha$.
We further note that $\tilde{X}_\alpha^\dag $ has the eigenvalue $\lambda^*_\alpha$. 
These conditions are fulfilled for $\lambda_\alpha = \rmi \omega_\alpha$ so that $\lambda_\alpha^* = - \lambda_\alpha$.
Now we can set
\beq
\tilde{X}_{1,2,3} = ( b_1, b_2,b_3) \,, ~~~~~~ \tilde{X}_{4,5,6} = ( b_1^\dag, b_2^\dag, b_3^\dag ) 
\eeq
where $\tilde{X}_{1-3}$ and $\tilde{X}_{4-6}$ have the eigenvalues $-\rmi \omega_{1-3}$ and $\rmi \omega_{1-3}$, respectively,
and the matrix $V$ turns out to be the matrix $T$ in Eq.~\eqref{eq:Tb}.
These frequencies are nothing but those in Eq.~\eqref{eq:HO_b}.
Now Eq.~\eqref{eq:HO_b} satisfies the time evolution,
\begin{align}
b_\alpha (t) 
& = \rme^{ -\rmi \omega_\alpha t } b_\alpha
= \rme^{ \rmi H_\perp^{\rm eff}  t } b_\alpha \rme^{ -\rmi H_\perp^{\rm eff}  t }
\,,
\notag \\
b_\alpha^\dag (t) 
& = \rme^{ \rmi \omega_\alpha t } b_\alpha^\dag
= \rme^{ \rmi H_\perp^{\rm eff}  t } b_\alpha^\dag \rme^{ -\rmi H_\perp^{\rm eff}  t }
\,,
\end{align}
as requested in Eq.~\eqref{eq:tilX_t_evo}.

%%%%%%%%%%%%%%%%%%%%%%
\subsubsection{Wavefunctions}
%%%%%%%%%%%%%%%%%%%%%%

To find the ground state wavefunctions for the transverse Hamiltonian, we look for the state such that
\beq
b_\alpha | 0_\perp \ra = T^{-1}_{\alpha i} X_i | 0_\perp \ra = 0 \,,~~~~~~(\alpha =1,2,3) \,.
\eeq
Writing  $x_\alpha = ( \bm{\rho}_\perp, R_y)$ and $p_\alpha = (\bm{p}_\rho, P_R^y)$,
this takes the form
\beq
\sum_{\beta=1}^3 \big( f_{\alpha \beta} x_\beta + g_{\alpha \beta} \, p_\beta  \big) | 0_\perp \ra = 0 \,,
\eeq
where $f_{\alpha \beta} = T^{-1}_{\alpha \beta} $ and $g_{\alpha \beta} = T^{-1}_{\alpha, \beta+3} $
are 3$\times$3 matrices.
The form of the solution is
\beq
\la \vx | 0_\perp \ra
= \calN_\perp \exp\bigg[ - \frac{\, 1 \,}{\, 2 \,} x_\alpha M_{\alpha \beta} x_\beta \bigg] \,,
\eeq
where the matrix $M(=M^T)$ can be computed as
\beq
M_{\alpha \beta} = \rmi \big[ g^{-1} f \big]_{\alpha \beta} 
\,,
\eeq
with the normalization constant is 
%$|\calN_\perp|^2 = \sqrt{\det M_R/\pi^3 }$ with $M_R \equiv {\rm Re} M$.
$|\calN_\perp |^2 = \sqrt{\det \calM/\pi^3 }$ where $\calM \equiv {\rm Re}M$.
The wavefunctions for excited states can be created by applying creation operators $b_\alpha^\dag$.

%%%%%%%%%%%%%%%%%%%%%%
\subsubsection{Matrix elements}
%%%%%%%%%%%%%%%%%%%%%%

In our computations of the matrix elements, 
we evaluate the expectation values of functions independent of $R_y$.
The electric potential is ($\calN_{z}^{\perp} = \calN_{n_z=0} \calN_\perp$)
% corrected
\begin{align}
\hspace{-0.2cm}
&\la V_E \ra_{\rm gs}
= |\calN_{z}^{\perp}|^2
\int_{R_y, \br} V_E (r)\,
\rme^{  - x_\alpha \calM_{\alpha \beta} x_\beta  - \frac{\, \Lambda^2 }{\, 2 \,} z^2  }
\notag \\
&= - \frac{\, 4 \alpha_s \,}{3} |\calN_{z}^{ \perp}|^2 \sqrt{ \frac{\, \pi \,}{\, \calM_{RR} \,} }
\int_{\br} 
\frac{\, \rme^{  - \rho^\perp_a \Sigma_{ab} \rho^\perp_b  - \frac{\, \Lambda^2 }{\, 2 \,} z^2  }  \,}{ r }
\,,
\end{align}
%
%% corrected
%\begin{align}
%\hspace{-0.2cm}
%\la V_E \ra_{\rm gs}
%&= |\calN_{z}^{\perp}|^2 |\calN_R|^2
%\int_{R_y, \br} V_E (r)\,
%\rme^{  - x_\alpha \calM_{\alpha \beta} x_\beta  - \frac{\, \Lambda^2 }{\, 2 \,} z^2  }
%\notag \\
%&= - \frac{\, 4 \alpha_s \,}{3} |\calN_{z}^{ \perp}|^2 
%\int_{\br} 
%\frac{\, \rme^{  - \rho^\perp_a \Sigma_{ab} \rho^\perp_b  - \frac{\, \Lambda^2 }{\, 2 \,} z^2  }  \,}{ r }
%\,,
%\end{align}
%%
where integrating $R_y$ renormalizes the matrix for $\bm{\rho}_\perp$,
\beq
\Sigma_{ab} = \calM_{ab} - \frac{\,  \calM_{Ra} \calM_{Rb} \,}{\, \calM_{RR} \,} \,,~~~(a,b=x,y) \,.
\eeq
%
%%
%\beq
%\Sigma_{ab} = M_{ab} - \frac{\, M_{Ra} M_{Rb} \,}{\, M_{RR} \,} \,,~~~~~~~(a,b=x,y)
%\eeq
%%
For the magnetic potential, we find (for $J={\rm I, II}$)
\begin{align}
\hspace{-0.2cm}
\la V_J^s \ra_{\rm gs}
&=  
\int_{\vq} V_J^s (\vq)\,
\rme^{  - \frac{1}{\, 4 \,} q^\perp_a (\Sigma^{-1})_{ab} q^\perp_b  - \frac{\, q_z^2 \,}{\, 2 \Lambda^2 \,}   }
\notag \\
&= 
\frac{\, 16 \pi  \alpha_s \,}{\, 3 \,}
\int_{\vq} 
 \rme^{  - q^\perp_a \tilde{ \Sigma }_{ab} q^\perp_b  }  \calF_s \calP_{J}
\,,
\end{align}
where $\calF_s$ and $\calP_J$ are given in Eqs.~\eqref{eq:calF_s} and \eqref{eq:calP}, respectively,
and $\tilde{\Sigma} \equiv \Sigma^{-1}/4 + I_{zz}/2\Lambda^2$.
Useful formulae to reduce the three dimensional integral over $(z, \rho_x, \rho_y)$
to one-dimensional integral is given in Appendix.~\ref{sec:int_formula}.

%%%%%%%%
\begin{figure}[t]
\vspace{-.1cm}
\begin{center}
\includegraphics[width=8.8 cm]{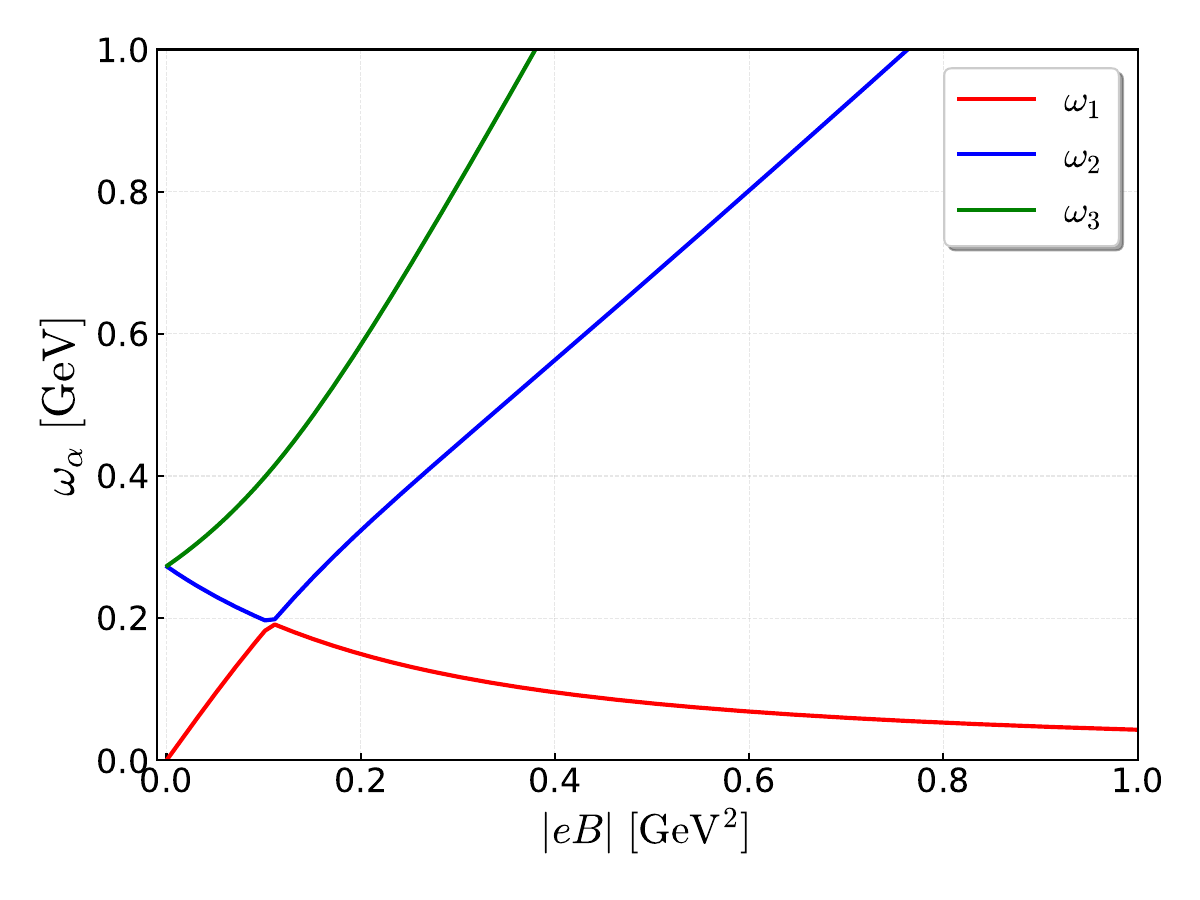}
\end{center}
\vspace{-.7cm}
\caption{Eigenfrequencies $\omega_\alpha$ ($\alpha=1,2,3$) for the transverse Hamiltonian $H_\perp^{\rm eff}$,
see Eq.~\eqref{eq:HO_b}.
The $u\bar{d}$ meson is considered.
These eigen-modes are the mixtures of the center-of-mass, $+$, and $-$ modes.
The level repulsion occurs around $|eB| \simeq 0.12\, {\rm GeV}^2$.
}
\label{fig:ud_omegas.pdf}
\end{figure}   
%%%%%%%%

%%%%%%%%
\begin{figure}[t]
\vspace{-.1cm}
\begin{center}
\includegraphics[width=8.8 cm]{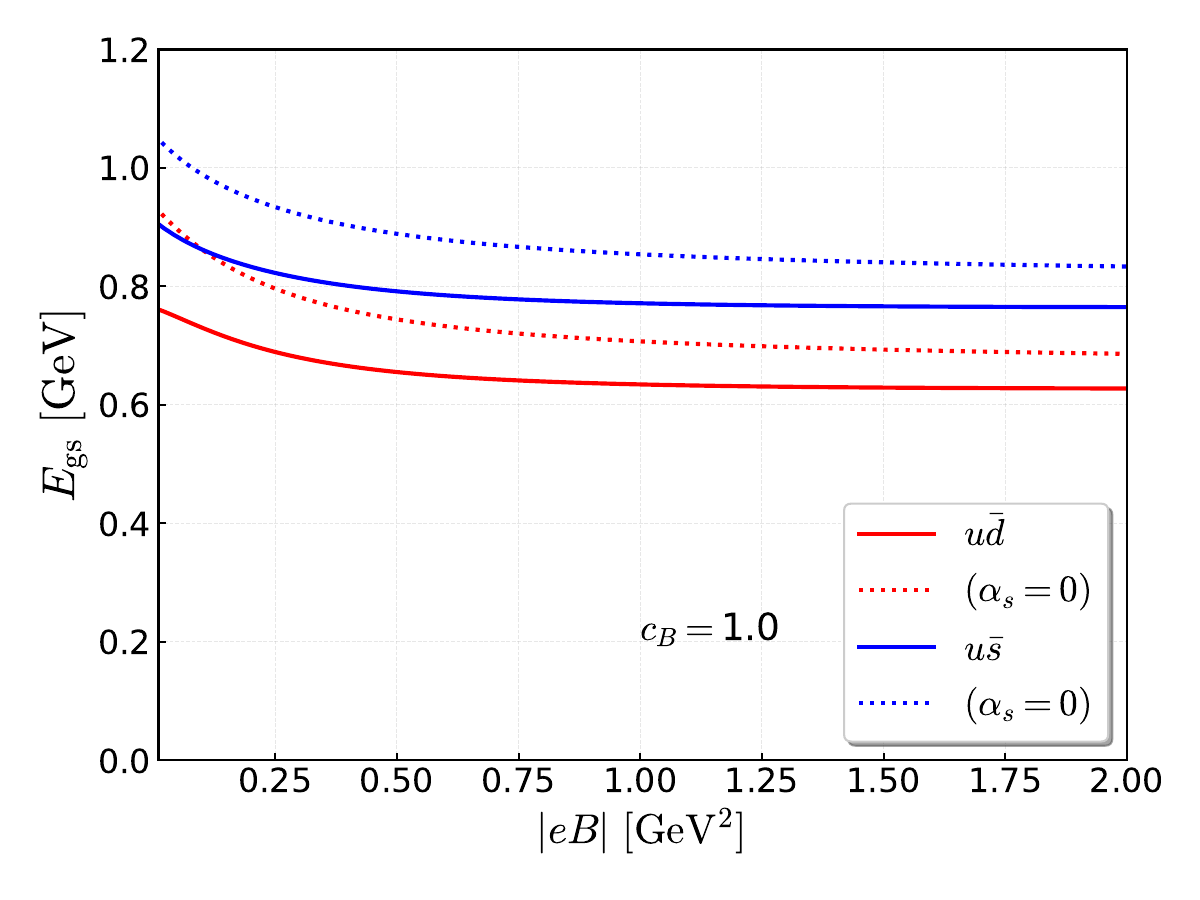}
\end{center}
\vspace{-.7cm}
\caption{Magnetic field dependence of the ground state energies for $u\bar{d}$ and $u\bar{s}$ mesons.
The solid and dotted lines represent calculations
 with and without short-range interactions ($\alpha_s (Q_B) \neq 0$ and $\alpha_s = 0$).
 For $Q_B$, we use the expression with $c_B =1.0$.
}
\label{fig:ud_us_alpha_on_off}
\end{figure}   
%%%%%%%%
%%%%%%%%
\begin{figure}[t]
\vspace{-.1cm}
\begin{center}
\includegraphics[width=8.8 cm]{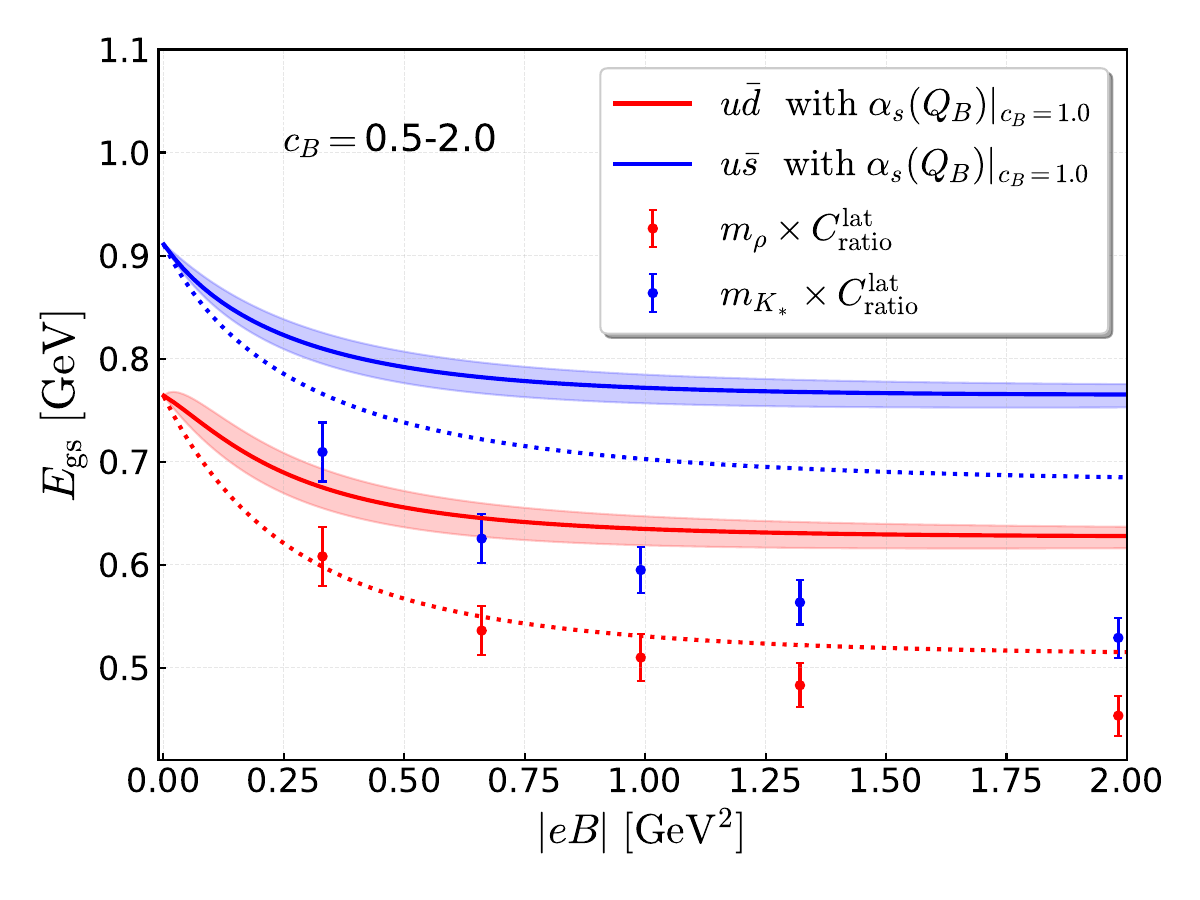}
\end{center}
\vspace{-.7cm}
\caption{Ground-state energy levels of $u\bar{d}$ and $u\bar{s}$ mesons as functions of $|eB|$,
calculated with the running coupling $\alpha_s (Q_B)$.
The colored bands represent the uncertainty originating from the scale parameter
$c_B \in [0.5, 2.0]$, where the solid lines denote the results for the central value, $c_B = 1.0$.
For comparison, the results for a constant coupling (corresponding to $c_B=0$) are indicated with the dotted curves.
We also plot the vacuum meson multiplied by the ratio $m_\rho (B)/m_\rho(0)$ measured on the lattice for $m_\pi=0.415$ GeV \cite{Bali:2017ian}.
}
\label{fig:ud_us_alpha_varying}
\end{figure}   
%%%%%%%%
%%%%%%%%
\begin{figure}[t]
\vspace{-.1cm}
\begin{center}
\includegraphics[width=8.8 cm]{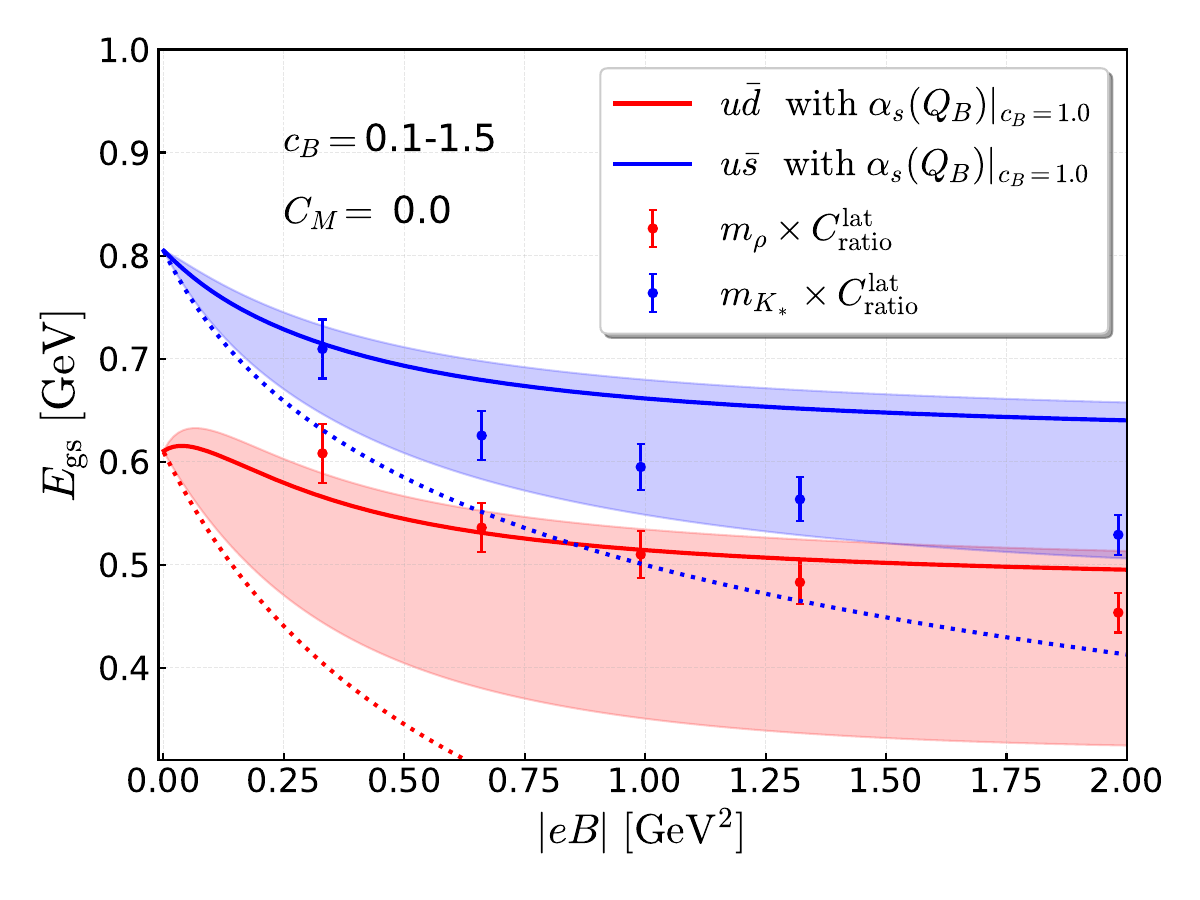}
\end{center}
\vspace{-.7cm}
\caption{Same as Fig.~\ref{fig:ud_us_alpha_varying},
but the color-magnetic interaction is switched off $(C_M=0)$.
The range of $c_B$ for the band is shifted to the lower values ($c_B = [0.1,1.5]$)
to explore the range for a better fit.
}
\label{fig:ud_us_alpha_varying_wo_CM}
\end{figure}   
%%%%%%%%

%%%%%%%%%%%%%%%%%%%%%%
\subsubsection{Spectra}
%%%%%%%%%%%%%%%%%%%%%%

Now we examine numerical results for the ground state spectra for $u\bar{d}$ and $u \bar{s}$ mesons.

We begin with the eigenfrequencies $\omega_\alpha$ ($\alpha=1,2,3$) of the transverse Hamiltonian $H_\perp^{\rm eff}$,
see Eq.~\eqref{eq:HO_b} and Fig.~\ref{fig:ud_omegas.pdf}.
In the limit of $e_\rho =0$ (see Sec.~\ref{sec:e_rho=0}),
we have already seen the center-of-mass motion with the energy cost $\sim |B|/M$,
and two orbital angular motions with the magnetic moments parallel and anti-parallel to the magnetic field,
with the energy cost of $\sim |B|/M$ and $\sim \lambda/|B|$, respectively.
The small energy in the latter is the consequence of the orbital Zeeman energy
that tends to cancel the zero-point energy of the orbital motion.

At $e_\rho \neq 0$, all these modes can mix,
but the large $B$ behaviors are similar to the trend found in the $e_\rho =0$ case;
we have two energetic modes ($\omega_2$ and $\omega_3$ modes) with the linear dependence on $B$,
and one low energy mode ($\omega_1$ mode) which decreases with $B$.
Extrapolating the assignment of quantum numbers for the $e_\rho =0$ case,
one can regard the $\omega_1$ mode to be dominated by the $\ell_z =+1$ excitation.
%
%Here it should be emphasized 
%that this $\ell_z=+1$ excitation is the mixture of angular excitations in center-of-mass and relative motions.
%The mixing with the center-of-mass motion allows hadronic states, which have the different parity in the rest frame at $B=0$, 
%to mix in the presence of magnetic fields.
%For instance, the vector $(\rho_+)_{s_z=1}^{\ell_z=0}$ and axial-vector $(a_{1+})^{\ell_z=0}_{s_z=1}$ mesons can mix.
%
Meanwhile, for small $B$, the lowest energy mode is dominated by the center-of-mass motion 
but it is eventually dominated by one of the orbital modes.
When these modes ($\omega_1$ and $\omega_2$) approach, 
the level repulsion occurs, see $\omega_1$ around $|eB| \sim 0.12\, {\rm GeV}^2$.

Shown in Fig.~\ref{fig:ud_us_alpha_on_off} 
is the magnetic field dependence of the ground state energies for $u\bar{d}$ and $u\bar{s}$ mesons,
with and without short-range interactions ($\alpha_s (Q_B) \neq 0$ and $\alpha_s = 0$).
For $Q_B$, we chose $c_B=1.0$.
For $\alpha_s=0$, as in the neutral meson cases,
the mass reduction occurs due to the effective disappearance of the transverse zero-point energy.
With the short-range interactions added in,
this trend does not change much.

In Fig.~\ref{fig:ud_us_alpha_varying}, we show 
the ground state energies for $u\bar{d}$ and $u\bar{s}$ mesons calculated with the running $\alpha_s(Q_B)$.
The colored bands represent the uncertainty originating from the scale parameter
$c_B \in [0.5, 2.0]$, where the solid lines denote the results for the intermediate value, $c_B = 1.0$.
The results for a constant $\alpha_s$ are also shown as dotted lines.
For comparison with lattice results, we plot the vacuum masses multiplied 
by the ratio $C_{\rm ratio}^{\rm lat} =m_{\rho_+}(B)/m_\rho(0)$ measured on the lattice for $m_\pi =0.415$ GeV \cite{Bali:2017ian}.
Since previous lattice analyses have shown that the pion mass dependence is mild between
$m_\pi \simeq 0.42$ GeV and $0.81$ GeV,
we applied the same ratio to the vacuum masses of $m_\rho$ and $m_{K*}$.

Compared with the lattice data, our model calculations seem to underestimate the mass reduction.
Increasing the value of $\alpha_s$, which affects both the color-electric (attractive) and magnetic (repulsive) interactions,
does not resolve the discrepancy.
One possible resolution is to assume that the color-magnetic interaction is weaker for $\rho$ and $K_{*+}$ 
at the relatively large current quark masses used in the lattice calculations \cite{Bali:2017ian}.
Indeed, if we switch off the color-magnetic interaction in our quark model (by setting $C_M=0$),
the loss of the repulsive term leads to a significant mass reduction, 
bringing the resulting mass ratio into better agreement with that obtained in the lattice calculation, 
as shown in Fig.~\ref{fig:ud_us_alpha_varying_wo_CM}.
At present, however,
we have not yet identified a definitive physical mechanism to explain the significant mass reduction observed on the lattice.
Further studies are called for.

%Compared to the neutral meson cases,
%the impacts of short-range interactions are substantially smaller,
%as seen from the smaller error band.
%The reason is that
%the typical distance between two guiding centers, $\bm{X} = \bm{X}_1 - \bm{X}_2$ (see Eq.~\eqref{eq:guiding}), 
%is larger than the neutral meson case where we could take $\bm{X}=0$ by choosing $\bm{K}_\perp = 0$.
%In the charged meson cases, the distance between two guiding centers fluctuates
%and the typical distance is effectively larger.

%%%%%%%%%%%%%%%%%%%%%%
\section{Summary} \label{sec:summary}
%%%%%%%%%%%%%%%%%%%%%%

We studied the properties of neutral and charged mesons
within a simple constituent quark model.
We first investigated how the confining potential 
affects the dynamics of charged particles in magnetic fields,
and then delineated the impact of the short-range interactions.
The qualitative trends in the mass reduction of neutral and charged mesons, as found in the lattice simulations, are reproduced.
The agreement with the lattice results can be improved to a semi-quantitative level
if a proper scale for the running coupling is chosen for the short-range interactions.

In addition to the ground-state energies, 
we also argue that the transverse kinetic terms  of neutral mesons have nontrivial behavior at finite $B$
which are substantially different from those obtained by extrapolating from a hadronic effective Lagrangian at $B=0$
\cite{Fukushima:2012kc,Hattori:2015aki,Kojo:2021gvm}.
The impact of transverse momenta is suppressed by a factor of $\sim 1/B$,
and hence at large $B$ there are many low-energy states.
This may be interpreted as an effective dimensional reduction of the meson dynamics.
Similar trends can be found for charged mesons with excitations of $\ell_z >0$.
In a hadron resonance gas with these low-lying mesons, 
thermal contributions at low temperature are significantly enhanced,
and this trend is consistent with the findings on the lattice \cite{Bali:2014kia}.

In this study we found that the details of the running of $\alpha_s$
are crucial for quantitative description of the hadronic spectra within the present framework,
especially for states with the spatially compact wavefunctions.
Unfortunately our choice of $Q^2$ remains largely heuristic and should be improved by more rigorous arguments \cite{Deur:2016tte}.
Also, quantitatively, our description of charged mesons are not satisfactory.
Considering the sensitivity of hadronic spectra to the choice of $\alpha_s$,
the $B$-dependence of the hadronic spectra 
should allow us to gain more insights on the running of $\alpha_s$ at the hadronic scale.

As mentioned in the Introduction,
this work is aimed to establish a reasonable (qualitative) baseline
to study multi-quark systems and hadronic matter in magnetic fields.
In subsequent papers, we discuss hadronic scattering at large $B$
and how such scattering processes can be used to describe multi-quark systems.
These results provide a useful framework for understanding
hadronic dynamics in strong magnetic fields.

%%%
\begin{acknowledgments}
We thank Gaoqing Cao and Heng-Tong Ding for insightful discussions,
and Dan Zhang, Gergely Endr{\H{o}}di, and Bastian Brandt for sharing their data with us.
This work is supported by
JSPS KAKENHI Grant No. 23K03377 and 26K07077 (TK);
and SOKENDAI Special Researcher Program (SI).

\end{acknowledgments}
%%%

%%%
\section*{Data Availability}
The lattice QCD data compared with our model results are available
in Ref.~\cite{Ding:2026qzu} for Fig.~\ref{fig:uu_ds_alpha_varying},
and
in Ref.~\cite{Bali:2017ian} for Figs.~\ref{fig:ud_us_alpha_varying}
and \ref{fig:ud_us_alpha_varying_wo_CM}.
The data from our model calculations are available from the authors
upon request.

\appendix

%%%%%%%%%%%%%%%%%%%
\section{Perturbative evaluation} \label{sec:pert_details}
%%%%%%%%%%%%%%%%%%%

To compute the corrections to the analytic expression in the $e_\rho=0$ (see Sec.~\ref{sec:pert_erho}),
we split the Hamiltonian by classifying terms by the order of $e_\rho$,
\beq
H_\perp^{\rm eff} = H^\perp_0 + H^\perp_{1} + H^\perp_{2} + \cdots \,.
\eeq
Below the reduced mass is expanded as
$\mu^{-1} = 4 M^{-1} + 4 \delta m^2 M^{-3} + \cdots$,
and $m_1^{-1} - m_2^{-1} = - 4 \delta m /M^2 + \cdots$.
Explicitly,
\begin{align}
H^\perp_0 
&=
 \frac{\, (P_R^y)^2 \,}{\, 2M \,}  
 + \frac{\, ( B e_R )^2 \,}{\, 2M \,} R_y^2
 \notag \\
 &
 + \frac{\, 2 (\vp_\rho^\perp)^2 \,}{\, M \,} 
+ \bigg[ \frac{\, (B e_R)^2 \,}{\, 32 M \,} + \lambda \, \bigg] \bm{\rho}_\perp^2  
- \frac{\, B e_R \,}{\, 2M \,} L_\rho^z 
\,,
\end{align}
is $O(e_\rho^0)$, and
\beq
\hspace{-0.7cm}
H^\perp_1
= 
 \frac{\, 2 B e_\rho \,}{\, M \,} R_y p_\rho^x
+ \frac{\, 5 B^2 e_R e_\rho \,}{\, 4M \,} \rho_y R_y 
- \frac{\, B e_\rho \,}{\, 2M \,} \rho_x p_R^y 
\,,
\eeq
is $O(e_\rho^1)$, and
\begin{align}
 H^\perp_2
&
= \frac{\, (B e_\rho)^2 \,}{\, 2 M \,}  R_y^2 + \frac{\, 2  \delta m^2 \,}{\, M^3 \,} (\vp_\rho^\perp)^2
\notag \\
& + \bigg[ \frac{ (B e_\rho)^2 }{\, 2 M \,} 
	+ \frac{\, B^2  e_R e_\rho \delta m\,}{\, 4 M^2 \,} 
	- \frac{\, (B e_R )^2 \delta m^2 \,}{\, 32 M^3 \,} 
	\bigg] \bm{\rho}_\perp^2	
	\notag \\
&
- \frac{\, 3 (B e_\rho)^2 \,}{\, 8 M \,} \rho_x^2
- \frac{\, 2 B e_\rho \delta m\,}{\, M^2 \,}  L_\rho^z
\,.
\end{align}
is regarded as $O(e_\rho^2)$.
Finally, we also split the Zeeman term as
\begin{align}
\hspace{-0.5cm}
H_{\rm Zm}
&= - \frac{\, e_1 B \,}{\, 2m_1 \,} \sigma_1^z - \frac{\, e_2 B \,}{\, 2m_2 \,} \sigma_2^z 
\notag \\
& 
= -  \frac{\, B e_R \,}{\, 2 M \,} \big( \sigma_1^z + \sigma_2^z \big)
+ \frac{\, B e_\rho \,}{\,  M \,}
	\big( \sigma_1^z - \sigma_2^z \big)	
	\notag \\
&~~~
 -  \frac{\, B e_\rho \delta m \,}{\,  M^2 \,}
	\big( \sigma_1^z + \sigma_2^z \big)	
+ O(\delta m^4)	
\,.	
\end{align}
%
%%
%\begin{align}
%\hspace{-0.5cm}
%H_{\rm Zm}
%&= - \frac{\, e_1 B \,}{\, 2m_1 \,} \sigma_1^z - \frac{\, e_2 B \,}{\, 2m_2 \,} \sigma_2^z 
%\notag \\
%& 
%= -  \frac{\, B e_R \,}{\, 2 M \,} \big( \sigma_1^z + \sigma_2^z \big)
%- \frac{\, B e_\rho \,}{\,  M \,}
%	\big( \sigma_1^z - \sigma_2^z \big)	
%	\notag \\
%&~~~
% -  \frac{\, B e_\rho \delta m \,}{\,  M^2 \,}
%	\big( \sigma_1^z + \sigma_2^z \big)	
%+ O(\delta m^4)	
%\,.	
%\end{align}
%%
where the first, second, and third terms correspond to 
$O(e_\rho^0)$, $O(e_\rho^1)$, and  $O(e_\rho^2)$, respectively.

For the transverse Hamiltonian of $O(e_\rho)$,
the first order perturbation is
\beq
\la H_1^\perp \ra = 0 \,,
\eeq
so that the corrections start from $O(e_\rho^2)$.
The $O(e_\rho^2)$ correction to the $n$-th level is as
\begin{align}
\big( E_{n_R}^{ \bm{n}_\perp } \big)_2
&= \sum_{\bm{n}'\neq \bm{n}} 
	\frac{\, \la \bm{n} | H_1^\perp | \bm{n}' \ra \la \bm{n}' | H_1^\perp | \bm{n} \ra \,}
		{\, \big( E_{n_R}^{ \bm{n}_\perp } \big)_0 -\big( E_{n_R'}^{ \bm{n}'_\perp } \big)_0 \,} 
+  \la \bm{n} | H_2^\perp | \bm{n} \ra 		
\notag \\
&= \delta_1 E_{n_R}^{ \bm{n}_\perp } 
+ \delta_2 E_{n_R}^{ \bm{n}_\perp }
+ \delta_{2, {\rm Zm} } E_{n_R}^{ \bm{n}_\perp }
\,.
\label{eq:pert_erho2_app}
\end{align}
To evaluate $\rho_x P_R^y$, $R_y p_\rho^x$, and $\rho_y R_y$ in $H_1^\perp$,
it is convenient to use the creation and annihilation operators
\beq
\hspace{-0.6cm}
R_y = \frac{\, 1 \,}{\,  \sqrt{2B e_R} \,} \big( b + b^\dag \big)
\,,~~
P_R^y = -\rmi \sqrt{ \frac{\,  B e_R \,}{\, 2 \,} } \big( b - b^\dag \big)
\,,
\eeq
and
\beq
\hspace{-0.5cm}
\rho_{i} = \frac{\, 1 \,}{\, \sqrt{2\kappa} \,}  \big( a_i + a_i^\dag \big)
\,,~~~
p_\rho^i = - \rmi \sqrt{ \frac{\, \kappa \,}{\, 2 \,} \,} \big( a_i - a_i^\dag \big)
\,,
\eeq
where $i = x,y$ and $\kappa =  B e_R \mu/2M$.

For perturbation treatments for $(\bm{\rho}_\perp, \bm{p}_\rho^\perp$), 
it is useful to label the unperturbed states by $(n_+, n_-)$ 
which is related to $(n_r, \ell_z)$ by $n_\pm = n_r + (|\ell_z| \pm \ell_z)/2$.
The first order Hamiltonian is ($\eta_1 = - \eta_2 = 1$)
\begin{align}
H_1^\perp
&= \rmi \frac{\, B e_\rho \,}{\, 2M \,} \sum_{j=\pm} 
\bigg[ \frac{\, \xi \,}{2} -  \frac{\, 5\eta_j \,}{\, 2 \xi \,} - \frac{1}{\, \xi \,}  \bigg]
a_j^\dag b^\dag 
\notag \\
&- \rmi \frac{\, B e_\rho \,}{\, 2M \,} \sum_{j=\pm} 
\bigg[ \frac{\, \xi \,}{2} - \frac{\, 5 \eta_j \,}{\, 2 \xi \,} + \frac{1}{\, \xi \,} \bigg]
a_j b^\dag 
+ ({\rm h.c.})
 \,,
\end{align}
where $\xi = \big[ 1 + 32M\lambda/(B e_R)^2 \big]^{1/4}$.

For the ground state, only $a_j^\dag b^\dag$ type terms yield
nonzero contributions.
The $a_+^\dag b^\dag$ increases the quanta $(n_+,n_-,n_R) \rightarrow (n_++1, n_-, n_R+1)$ 
with the excitation energy
\beq
E^{\Delta n_+=1 }_{\Delta n_R =1 }
= \frac{\, B e_R  \,}{\, M \,} 
+ \frac{\, \calB_R - B e_R \,}{\, 2M \,} 
\,,
\eeq
while the $a_-^\dag b^\dag$ increases the quanta $(n_+,n_-,n_R) \rightarrow (n_+, n_- +1, n_R+1)$ 
with the energy
\beq
E^{\Delta n_-=1 }_{\Delta n_R =1 }
= \frac{\, B e_R  \,}{\, M \,} 
+ \frac{\, \calB_R + B e_R \,}{\, 2M \,} 
\,,
\eeq
Hence the energy shift $\delta_1 E$ for the ground state is
\begin{align}
\hspace{-0.1cm}
 \delta_1 E_{\rm gs} 
 = - \frac{\, ( B e_\rho )^2 \,}{\, 2M \,} 
\bigg[
\frac{\, \big( \frac{\, \xi \,}{2} - \frac{\, 7 \,}{\, 2 \xi \,}   \big)^2 \,}
	{\,  \calB_R + B e_R \,}
+
\frac{\, \big(  \frac{\, \xi \,}{2} + \frac{\, 3 \,}{\, 2 \xi \,}  \big)^2 \,}
	{\,  \calB_R + 3 B e_R \,}
\bigg]		
\,.
\end{align}
Next we evaluate $\delta_2 E$.
The straightforward calculations yield
\begin{align}
\big\la \bm{n}_\perp |  \bm{\rho}_\perp^2 | \bm{n}_\perp \big\ra
&
= \frac{\, 8 \,} {\,  B e_R \xi^2 \,}\,
	\big( n_+ + n_- + 1 \big) \,,
\notag \\
\big\la \bm{n}_\perp |  \big( \bm{p}_\rho^\perp \big)^2 | \bm{n}_\perp \big\ra
&
= \frac{\,  B e_R \xi^2  \,}{\, 8 \,} \,
	\big( n_+ + n_- + 1 \big) \,,
\notag \\
\big\la n_R | R_y^2 | n_R \big\ra
&= 	
\frac{\, 1 \,}{\,  2B e_R \,} \big( 2n_R + 1 \big) \,.
\end{align}
The corresponding energy shift is
\begin{align}
\hspace{-0.2cm}
 \delta_2 E_{\rm gs} 
 = \frac{\, B e_\rho^2  \,}{\, 4M e_R \,} %\frac{\, e_\rho^2 \,}{\, e_R \,} 
 	\frac{\, \xi^2 + 10 \,} {\, \xi^2 \,}
+ \frac{\, 2 B e_\rho \delta m  \,}{\, M^2 \xi^2 \,} 	
+ \frac{\, 8 \delta m^2 \lambda  \,}{\, M^2 B e_R \xi^2 \,} 	
\,,
\end{align}
where we used $\xi^4-1 = 32 M\lambda/(B e_R)^2$.
Including the corrections to the Zeeman energy, the results are summarized as
\begin{align}
\hspace{-0.1cm}
 &\delta_1 E_{\rm gs} 
  = - \frac{\, ( B e_\rho )^2 \,}{\, 2M \,} 
\bigg[
\frac{\, \big( \frac{\, \xi \,}{2} - \frac{\, 7 \,}{\, 2 \xi \,}   \big)^2 \,}
	{\,  \calB_R + B e_R \,}
+
\frac{\, \big(  \frac{\, \xi \,}{2} + \frac{\, 3 \,}{\, 2 \xi \,}  \big)^2 \,}
	{\,  \calB_R + 3 B e_R \,}
\bigg]		
\,,
\notag \\
\hspace{-0.1cm}
&  \delta_2 E_{\rm gs} 
 = \frac{\, B e_\rho^2  \,}{\, 4M e_R \,} %\frac{\, e_\rho^2 \,}{\, e_R \,} 
 	\frac{\, \xi^2 + 10 \,} {\, \xi^2 \,}
+ \frac{\, 2 B e_\rho \delta m  \,}{\, M^2 \xi^2 \,} 	
+ \frac{\, 8 \delta m^2 \lambda  \,}{\, M^2 B e_R \xi^2 \,} 
\,,
\notag \\
%&\delta_2 E_{\rm gs} 
% = \frac{\, B e_\rho^2  \,}{\, 4M e_R \,} %\frac{\, e_\rho^2 \,}{\, e_R \,} 
% 	\frac{\, \xi^2 + 10 \,} {\, \xi^2 \,}
%+ \frac{\, 2 B e_\rho \delta m  \,}{\, M^2 \xi^2 \,} 	
%+ \frac{\, 4 \delta m^2 \lambda  \,}{\, M^2 B e_R \,} 	
%\,, 
%\notag \\
&\delta_{\rm 2, Zm} E_{\rm gs} 
=
- \frac{\, 2 B e_\rho \delta m  \,}{\, M^2 \xi^2 \,} 	
\,,
\label{eq:ch_gs_2}
\end{align}
where $\xi = \big[ 1 + 32 M\lambda/(B e_R)^2 \big]^{1/4}$
and we have used $\xi^4-1 = 32 M\lambda/(B e_R)^2$
in computations of $\delta_2 E_{\rm gs} $.

In the limit of $B\rightarrow \infty$, the scaling of parameters is
$\calB_R \rightarrow B e_R$, $\xi \rightarrow 1$.
The perturbative corrections of $O(Be_\rho^2)$
are assembled to cancel,
\begin{align}
\delta_1 E_{\rm gs} 
&~\rightarrow~
- \frac{\, 11 B e_\rho^2 \,}{\, 4M e_R \,} \,,
\notag\\
\delta_2 E_{\rm gs} 
&~\rightarrow ~
\frac{\, 11 B e_\rho^2 \,}{\, 4M e_R \,} 
+ \frac{\, 2 B e_\rho \delta m  \,}{\, M^2 \,} 	
\,,
\notag \\
\delta_{\rm 2, Zm} E_{\rm gs} 
&~\rightarrow~
- \frac{\, 2 B e_\rho \delta m  \,}{\, M^2 \,} 	
\end{align}
and hence the $B$-dependence appears only through $O(M\lambda/B^2)$.

%%%%%%%%%%%%%%%%%%%
\section{Formulae for matrix elements} \label{sec:int_formula}
%%%%%%%%%%%%%%%%%%%

For perturbative evaluation of short-range potentials for neutral and charged mesons,
here we summarize some useful formulae.
We begin with
\beq
\frac{1}{\, \sqrt{ \xi^2 + a^2 } \,}
= %\frac{1}{\, \sqrt{ \pi } \,} 
\int_0^\infty \rmd s \, \frac{\, \rme^{-s(\xi^2 + a^2)} \,}{\, \sqrt{\pi s} \,}
\,.
\eeq
In computations of $V_E$, the three-dimensional integral is reduced to the one-dimensional integral,
\begin{align}
I_E [\Sigma]
&=
\int_{\br} 
\frac{\, \rme^{  - \bm{x}^T \Sigma \bm{x}  }  \,}{ |\bm{x}-\bm{x}_0| }
\notag \\
&
= % \frac{1}{\, \sqrt{ \pi } \,}
 \int_0^\infty \frac{\, \rmd s \,}{\, \sqrt{\pi s} \,} \int_{\br}
\rme^{  -  \bm{x}^T \Sigma \bm{x} - s (\bm{x} - \bm{x}_0)^2  }
\notag \\
&
=  \int_0^\infty \rmd s ~
\frac{ \pi \rme^{ s^2 \bm{x}_0^T (\Sigma + s I)^{-1} \bm{x}_0 - s \bm{x}_0^2 } }{\, \sqrt{ s \det\big( \Sigma + s I \big) }  \,}  
\,.
\end{align}
For the magnetic potential, we encounter the following types of integrals
\begin{align}
I^M_0 [\tilde{\Sigma}]
&= \int_{\vq} \rme^{ - \bm{q} \tilde{\Sigma} \bm{q} -\rmi \bm{q}_\perp \cdot \bm{r}_\perp^0 }
=  \frac{1}{\, 8 \pi^{3/2} \,} \frac{ w (\bm{r}_\perp^0)  }{\, \sqrt{ \det \tilde{\Sigma} }  \,}  \,,
%\notag 
\\
\tilde{I}^M_0 [A]
& = \int_{\vq}\frac{\, \rme^{ - q_a \tilde{\Sigma}_{ab} q_b -\rmi \bm{q}_\perp \cdot \bm{r}_\perp^0 } \,}{\, \bm{q}^2 + m^2 \,}
\notag \\
& = \frac{1}{\, 8 \pi^{3/2} \,}
\int_0^\infty \rmd s \, \frac{ \rme^{-sm^2  } w (\bm{r}_\perp^0)  }{\, \sqrt{ \det\big( \tilde{\Sigma} + s I \big) }  \,}  
 \,.
\end{align}
where 
\beq
w (\bm{r}_\perp^0) =  \rme^{ - \frac{1}{\, 4 \,} ( \bm{r}_\perp^{0} )^T \tilde{\Sigma}^{-1} \bm{r}_\perp^0  } \,.
\eeq
We also compute the integral
\begin{align}
& I^M_j [\tilde{\Sigma}]
=
\int_{\vq} \rme^{ - \bm{q}^T \tilde{\Sigma} \bm{q} -\rmi \bm{q}_\perp \cdot \bm{r}_\perp^0 } \, \frac{\, q_j^2 \,}{\, \bm{q}^2 \,}
\notag \\
& = - \frac{\partial}{\, \partial \tilde{\Sigma}_{jj} \,}
\int_0^\infty \rmd s \,\int_{\vq} \rme^{ - \bm{q}^T \tilde{\Sigma} \bm{q} - s \bm{q}^2 -\rmi \bm{q}_\perp \cdot \bm{r}_\perp^0 } 
\notag \\
& = - \frac{\partial}{\, \partial \tilde{\Sigma}_{jj} \,} \frac{1}{\, 8 \pi^{3/2} \,}
\int_0^\infty \rmd s \, \frac{\, w (\bm{r}_\perp^0)  \,}{\, \sqrt{ \det\big( \tilde{\Sigma}+ s I \big) }  \,}  
 \,,
\end{align}
and
\begin{align}
&\tilde{I}^M_j [\tilde{\Sigma}]
=
\int_{\vq} \rme^{ - \bm{q}^T \tilde{\Sigma} \bm{q} -\rmi \bm{q}_\perp \cdot \bm{r}_\perp^0 } \, \frac{\, 1 \,}{\, \bm{q}^2 + m^2\,}  \frac{\, q_j^2 \,}{\, \bm{q}^2 \,} 
\notag \\
& = - \frac{\partial}{\, \partial A_{jj} \,} \!
\int_0^\infty\!\! \rmd u \int_0^\infty \!\! \rmd t \,\int_{\vq} \rme^{ - \bm{q}^T \tilde{\Sigma} \bm{q} - t (\bm{q}^2 +m^2) - u \bm{q}^2 -\rmi \bm{q}_\perp \cdot \bm{r}_\perp^0 } 
\notag \\
& = - \frac{\partial}{\, \partial A_{jj} \,} \!
\int_0^\infty \rmd s \int_0^s \rmd t \,\int_{\vq} \rme^{ - \bm{q}^T [  \tilde{\Sigma}+ s I ] \bm{q} - t m^2 -\rmi \bm{q}_\perp \cdot \bm{r}_\perp^0 } 
\notag \\
& = - \frac{\partial}{\, \partial A_{jj} \,} \frac{1}{\, 8 \pi^{3/2} \,}
\int_0^\infty \!\! \rmd s \, \frac{ ( 1 - \rme^{- s m^2} ) w (\bm{r}_\perp^0)  \,}{\, m^2 \sqrt{ \det\big( \tilde{\Sigma} + s I \big) }  \,}  
 \,.
\end{align}
In particular, when $A$ is diagonal,
% corrected
\begin{align}
\hspace{-0.2cm}
I^M_j [\tilde{\Sigma}]
& = \frac{\, w (\bm{r}_\perp^0) \,}{\, 16 \pi^{3/2} \,}
\int_0^\infty \rmd s
 \frac{\, 1 \,}{\, \tilde{\Sigma}_{jj} + s \,} \prod_k \frac{ 1 }{\, \sqrt{ \tilde{\Sigma}_{kk} + s }  \,}  
 \,,
%\notag 
\\
\tilde{I}^M_j [\tilde{\Sigma}]
& = \frac{ w (\bm{r}_\perp^0)  }{\, 16 \pi^{3/2} \,}
\int_0^\infty \rmd s \,
 \frac{\, 1 - \rme^{- s m^2}  \,}{\, m^2 \big( \tilde{\Sigma}_{jj} + s \big) \,} 
 \prod_k \frac{ 1 }{\, \sqrt{ \tilde{\Sigma}_{kk} + s }  \,} 
 \,.
 \notag \\
%& \hspace{3.0cm} 
% \times \prod_k \frac{ 1 }{\, \sqrt{ \tilde{\Sigma}_{kk} + s }  \,}  
\end{align}
With these formulae and setting $ \tilde{\Sigma}' \equiv \tilde{\Sigma} + \mu_{\rm IR}^{-2} I$,
we compute the integrals appearing in the magnetic energy as
\begin{align}
\calG_0
& \equiv \int_{\vq} 
\rme^{ - \vq^T \tilde{\Sigma} \vq -\rmi \bm{q}_\perp \cdot \bm{r}_\perp^0 }  \calF_s
\notag \\
& = \frac{\, I^M_0 [ \tilde{\Sigma}' ] \,}{\, m_1 m_2 \,} 
 + \tilde{I}^M_0 [ \tilde{\Sigma}]  - \tilde{I}^M_0 [ \tilde{\Sigma}' ] 
 \,,
 %\notag 
 \\
\calG_j
& \equiv \int_{\vq} 
\rme^{ - \vq^T \tilde{\Sigma} \vq -\rmi \bm{q}_\perp \cdot \bm{r}_\perp^0 }  \calF_s  \frac{\, q_j^2 \,}{\, \bm{q}^2 \,} 
 \notag \\
& = \frac{\, I^M_j [ \tilde{\Sigma}' ] \,}{\, m_1 m_2 \,} 
 + \tilde{I}^M_j [ \tilde{\Sigma} ]  - \tilde{I}^M_j [ \tilde{\Sigma}' ] 
 \,.
\end{align}
The magnetic energy can be written as
\begin{align}
 \la V_{\rm I}^s \ra
& = \frac{\, 16 \pi  \alpha_s \,}{\, 3 \,}
\int_{\vq} 
\rme^{ - \vq^T \tilde{\Sigma} \vq -\rmi \bm{q}_\perp \cdot \bm{r}_\perp^0 } 
  \frac{\, \calF_s \,}{\, 2 \,} \bigg( 1 + \frac{\, q_z^2 \,}{\, \vq^2 \,} \bigg)
\notag 
\\
& 
= \frac{\, 8 \pi  \alpha_s \,}{\, 3 \,} \big( \calG_0 + \calG_z \big)
 \,,
 %\notag
  \\
 \la V_{\rm II}^s \ra
& = \frac{\, 16 \pi  \alpha_s \,}{\, 3 \,}
\int_{\vq} 
\rme^{ - \vq^T \tilde{\Sigma} \vq -\rmi \bm{q}_\perp \cdot \bm{r}_\perp^0  } 
  \frac{\, \calF_s \,}{\, 2 \,} \bigg( 1 - \frac{\, 3 q_z^2 \,}{\, \vq^2 \,} \bigg)
\notag \\
& 
= \frac{\, 8 \pi  \alpha_s \,}{\, 3 \,} \big( \calG_0 - 3 \calG_z \big)
 \,.
\end{align}

%
%%%%%%%%%%%%%%%%%%%%
%\section{The $\omega_1$ excitations in charged mesons} \label{sec:omega1_charged}
%%%%%%%%%%%%%%%%%%%%
%
%We consider the $\omega_1$ excitations in charged mesons.
%%
%\beq
%b_1^\dag| 0 \ra 
%= \sum_{\beta=1}^3 \big( f_{1 \beta}^* x_\beta + g_{1 \beta} \, p_\beta  \big) | 0 \ra
%\eeq
%%
%%
%\beq
%\sum_{\beta=1}^3 \big( f_{\alpha \beta} x_\beta + g_{\alpha \beta} \, p_\beta  \big) | 0_\perp \ra = 0 \,,
%\eeq
%%
%where $f_{\alpha \beta} = T^{-1}_{\alpha \beta} $ and $g_{\alpha \beta} = T^{-1}_{\alpha, \beta+3} $
%are 3$\times$3 matrices.
%The form of the solution is
%%
%\beq
%\la \vx | 0_\perp \ra
%= \calN_\perp \exp\bigg[ - \frac{\, 1 \,}{\, 2 \,} x_\alpha M_{\alpha \beta} x_\beta \bigg] \,,
%\eeq
%%
%where the matrix $M$ can be computed as
%%
%\beq
%M_{\alpha \beta} = - \rmi \big[ g^{-1} f \big]_{\alpha \beta} 
%\,,
%\eeq
%%

\bibliography{ref}

\end{document}